\documentclass[prd,aps,showpacs,superscriptaddress,nofootinbib,amsmath,amssymb]{revtex4}
\usepackage{graphicx}
\usepackage{amstext,amsfonts,bbm}
\usepackage{bm}
\def\slashchar#1{\setbox0=\hbox{$#1$}
   \dimen0=\wd0 \setbox1=\hbox{/} \dimen1=\wd1
   \ifdim\dimen0>\dimen1 \rlap{\hbox to \dimen0{\hfil/\hfil}} #1
   \else  \rlap{\hbox to \dimen1{\hfil$#1$\hfil}} / \fi}

\begin{document}


\title{Weak charged and neutral current induced one pion production off the nucleon}
\author{M. Rafi Alam}
\author{M. Sajjad Athar}
\author{S. Chauhan\footnote{Corresponding author: niharikavatsa21@gmail.com}}
\author{S. K. Singh}
\affiliation{Department of Physics, Aligarh Muslim University, Aligarh - 202 002, India}

\begin{abstract}
 We present a study of neutrino/antineutrino induced charged and neutral 
 current single pion production off the nucleon. For this, 
 we have considered  $P_{33}(1232)$ resonance, non-resonant background terms,
 other higher resonances like $P_{11}(1440)$, $S_{11}(1535)$, $D_{13}(1520)$, $S_{11}(1650)$ and $P_{13}(1720)$.
     For the non-resonant background terms a microscopic approach based on SU(2) non-linear sigma model has been used.
       The vector form factors for the resonances are obtained  
     by using the relationship between the electromagnetic resonance form factors 
     and helicity amplitudes provided by MAID. 
     Axial coupling $C_5^{A}(0)$ in the case of $P_{33}(1232)$ resonance is obtained by  fitting 
     the ANL and BNL $\nu$-deuteron reanalyzed scattering  data. The results are presented with 
     and without deuteron effect for the total scattering cross sections for 
     all possible channels viz. $\nu_l(\bar\nu_l)~+~N\rightarrow l^-(l^+)~+~N^\prime~+~\pi^i$; 
     $\nu_l(\bar\nu_l)~+~N\rightarrow \nu_l(\bar\nu_l)~+~N^\prime~+~\pi^i$, 
     where $N, N^\prime=p,n$, $\pi^i=~\pi^\pm$ or $\pi^0$ and $l=e, \mu$.
   \end{abstract}
\maketitle
\section{Introduction}
A precise knowledge of (anti)neutrino-nucleus cross sections is an important input in minimizing
the systematic errors  in the analysis of (anti)neutrino oscillation experiments.
Most of these experiments are presently being done in  the (anti)neutrino energy region of a few GeV. 
In this energy region of accelerator experiments like 
MiniBooNE~\cite{AguilarArevalo:2009ww,AguilarArevalo:2010bm,AguilarArevalo:2010xt,Katori:2013nca}, 
T2K~\cite{Abe:2014tzr,Abe:2013xua}, NO$\nu$A~\cite{Jediny:2014lda}, 
MicroBooNE~\cite{Chen:2007ae}, ArgoNeuT~\cite{Acciarri:2014eit,Spitz:2011wba}, LBNO~\cite{Patzak:2012rz},
MINOS~\cite{Saakian:2004cf}, 
DUNE~\cite{dune}, etc. major 
contribution to the neutrino-nucleus cross section 
comes from the quasielastic process and the inelastic process of single pion production(SPP). 
The basic reaction mechanism for quasielastic process is widely studied in the Standard Model
and there exist many calculations of nuclear medium effects(NME) using various nuclear models
 which have been summarized recently in several review articles~\cite{Benhar:2015wva,Alvarez-Ruso:2014bla,
 Gallagher:2011zza,Morfin:2012kn,Formaggio:2013kya}. In the case of single pion production processes 
 from nuclear target, nuclear medium effects 
 play an important role in the production process. In addition to this, the produced pion, being a hadron interacts with the 
 residual nucleus. These pions may get absorbed in the nucleus or may change their charge state through charge exchange 
 pion nucleon scattering processes like $\pi^-~+~p \rightarrow \pi^0~+~n$, 
$\pi^0~+~p \rightarrow \pi^+~+~n$, etc. Therefore, the final state interaction(FSI) effect of the pion with the nucleus 
 has also to be taken into account. Moreover, presently there is lack of consensus on theoretical 
modeling of basic reaction mechanism of (anti)neutrino 
induced single pion production from free nucleons. 

At neutrino energies of $\sim$ 1 GeV, the single pion  production channels make
a significant contribution to the cross section for charged  lepton production
and are important processes to be considered in the analysis of 
oscillation  experiments which select charged current inclusive events as signal. 
 In experiments which select the quasielastic 
production of  charged leptons as  signals for the analysis of oscillation experiments, 
single pion production channel gives rise to background contribution.
For example,  neutral current induced neutral pion  production is a background to 
$\nu_e -$appearance oscillation
experiments while charged current events producing charged pions contribute to 
background in $\nu_\mu -$disappearance experiments.
It is, therefore, very important to theoretically  understand and model the single
pion production processes on  nuclear targets like
$^{12}C$, $^{16}O$,  $^{40}Ar$, $^{56}Fe$, $^{208}Pb$, which 
are being used in the present experiments. 

The present attempts to explain the experimental data on weak pion
production in neutrino/antineutrino reactions from nucleons bound inside 
 the nucleus like the experiments performed at 
 MiniBooNE~\cite{AguilarArevalo:2009ww,AguilarArevalo:2010bm,AguilarArevalo:2010xt,Katori:2013nca}, 
SciBooNE\cite{Katori:2013nca}, K2K\cite{Gran:2011zz, Mariani:2009zzb, Mariani:2010ez, Whitehead:2008zza} 
 and more recently from MINER$\nu$A 
 collaboration\cite{Eberly:2014mra, Fiorentini:2013ezn, Fields:2013zhk, Walton:2014esl, Tice:2014pgu} 
 have highlighted the 
inadequacy of our present 
 understanding of nuclear medium and final state interaction effects. The experimental results of single pion production  
and their comparison with the various 
theoretical calculations have also 
necessitated the need to re-examine the basic reaction mechanism for the production of single pion
from free nucleon target. 
 
  The various possible reactions which may contribute to the single pion production 
  either through charged current or neutral current neutrino/antineutrino induced reaction on a nucleon target
  are the following:
  
  Charged current(CC) induced processes:
\begin{align}
 \nu_{_l} p &\to l^- p \pi^+  &  \bar \nu_{_l} n &\to l^+ n \pi^- \nonumber \\
 \nu_{_l} n &\to l^- n \pi^+  &  \bar \nu_{_l} p &\to l^+ p \pi^- \nonumber \\
  \nu_{_l} n &\to l^- p \pi^0  &  \bar \nu_{_l} p &\to l^+ n \pi^0 \quad \qquad ; \; l = e,\mu
\label{eq:CC_all}
\end{align}
and neutral current(NC) induced processes:
 \begin{align}
 \nu_{_l} p &\to \nu_{_l} n \pi^+  &  \bar \nu_{_l} p &\to \bar \nu_{_l} p \pi^0 \nonumber \\
 \nu_{_l} p &\to \nu_{_l} p \pi^0  &  \bar \nu_{_l} p &\to \bar \nu_{_l} n \pi^+ \nonumber \\
  \nu_{_l} n &\to \nu_{_l} n \pi^0  &  \bar \nu_{_l} n &\to \bar \nu_{_l} n \pi^0 \nonumber \\
   \nu_{_l} n &\to \nu_{_l} p \pi^-  &  \bar \nu_{_l} n &\to \bar \nu_{_l} p \pi^-.
  \label{eq:NC_all}
\end{align}
The existing experimental data on single pion production process from
(almost)free  nucleons are available only from the 
old bubble chamber experiments performed at ANL~\cite{Radecky:1981fn} and BNL~\cite{Kitagaki:1986ct}
from deuteron/hydrogen targets.
These data on $\nu_{\mu} p \to \mu^- p \pi^+$ differ with each other by about $30-40\%$ which 
has been attributed in the past to different flux normalization in these experiments. 
These data, when used to fix various parameters of theoretical models of reaction 
mechanisms for single pion production, give rise to considerable uncertainties in the 
determination of these parameters, which in turn lead to higher uncertainties in predicting  
the single pion production  cross section  from nuclear targets.
 Recently reanalysis of the old bubble chamber data by the two independent 
 groups have tried to arrive at a consistent set of data 
 from ANL~\cite{Radecky:1981fn} and BNL~\cite{Kitagaki:1986ct} experiments either 
 by minimizing the neutrino flux uncertainties~\cite{Graczyk:2009qm,Graczyk:2014dpa}
 or by reconstructing the data using the cross section ratio for single pion production to the 
 quasielastic processes,  and observed quasielastic cross sections~\cite{Wilkinson:2014yfa}.
It is hoped that the use of reanalyzed/reconstructed data on free nucleon targets will help 
towards a better understanding of pion production reaction mechanism. Furthermore, using these data a better  
determination of various parameters to be used in the theoretical calculations may also be possible. 

Theoretically, the weak single pion production has been studied for almost 50 years. The early calculations were
 based on dynamical models using dispersion theory or quark models or an effective Lagrangian field 
theory~\cite{Berman:1965iu,Bell:1996mw,Bell:1970mc,Albright:13.673,Albright:14.324,Albright:1965zz} 
and a comprehensive summary has been given by C. H. Llewellyn Smith~\cite{LlewellynSmith:1971zm}.
Since then calculations have been made either by using quark models~
\cite{Schreiner:1973mj,Ravndal:1972ws,Rein:1980wg,Kuzmin:2003ji,Kuzmin:2004ya,Wu:2013kla} 
or dynamical models~\cite{Sato:2003rq,Wu:2014rga,Kamano:2013iva},
but most of the recent calculations have been done using effective Lagrangian field theory
\cite{SajjadAthar:2009rc,SajjadAthar:2009rd,Athar:2005hu,Leitner:2006ww,Hernandez:2007qq,Lalakulich:2010ss}, 
where the calculations are performed using a $\Delta(1232)$ dominance model 
to successfully explain 
the experimental data on $\nu_\mu p \to \mu^-  p  \pi^+$ channel by fitting
the $N-\Delta$ transition form factors. These form factors are found to be consistent  with the  predictions of the hypothesis
of conserved vector current(CVC) and partial conservation of axial vector current(PCAC).
However, these analyses have large uncertainties mainly due to the incompatibility of single pion
production data in this channel from ANL~\cite{Radecky:1981fn} and BNL~\cite{Kitagaki:1986ct} 
experiments which also reflect into the 
determination of $N-\Delta$ transition form factors. 
These uncertainties are further enhanced as systematic errors due to nuclear medium 
and FSI effects, which also come into play, when applied to 
explain the data from nuclear targets like $^{12}C$, $^{16}O$ and other heavier nuclear targets. 

The recent tension between the experimental results on pion production from $^{12}C$ nuclear target in the
MiniBooNE~\cite{AguilarArevalo:2009ww,AguilarArevalo:2010bm,AguilarArevalo:2010xt,Katori:2013nca} and MINER$\nu$A 
 ~\cite{Eberly:2014mra, Fiorentini:2013ezn, Fields:2013zhk, Walton:2014esl, Tice:2014pgu} experiments
 have also highlighted the inadequacy of our present theoretical 
 understanding of single pion production processes from nuclear targets.
 One of the main concerns in the theoretical modeling of basic reaction mechanism for SPP is 
 the role of non-resonant background terms and  the
 contribution of higher resonances beyond
 the $\Delta(1232)$ dominance model.
 
The role of non-resonant background terms was also emphasized in earlier analyses of experimental data 
from ANL~\cite{Radecky:1981fn} and BNL~\cite{Kitagaki:1986ct} experiments. It was concluded that 
there might be a sizable contribution from the non-resonant background terms coming  
specially in neutrino-neutron channels like $\nu_\mu  n \to \mu^-  p  \pi^0$  and 
$\nu_\mu  n \to \mu^-  n  \pi^+$, even though they may be small in $\nu_\mu  p \to \mu^-  p  \pi^+$ channel.
Indeed a theoretical calculation by Fogli and Nardulli\cite{Fogli:1979cz,Fogli:1979qj} in an effective 
Lagrangian field theoretical model 
has shown that the inclusion of pion pole, nucleon pole, and other $I=\frac12$ resonance contribution 
leads to a better explanation of
ANL~\cite{Radecky:1981fn}, BNL~\cite{Kitagaki:1986ct}
and CERN~\cite{Lee:1976wr,Bell:1978qu,Allen:1980ti} data. 

In recent times, the need for inclusion of non-resonant and resonant ($I=\frac12$ channel) background 
terms to the dominant $\Delta(1232)$ contributions to explain the neutrino induced single pion 
production has been emphasized by many 
authors\cite{Hernandez:2007qq,Zhang:2012aka,Lalakulich:2012cj,Lalakulich:2013iaa,Leitner:2008ue} and numerical 
calculations have been performed to explain the present data.
These calculations have been done either phenomenologically~\cite{Lalakulich:2012cj,Lalakulich:2013iaa,Leitner:2008ue} or in an effective 
Lagrangian field theoretical model\cite{Hernandez:2007qq,Zhang:2012aka}. 
However, there is no consensus on the treatment of background terms 
whether they should be added coherently or incoherently to the dominant $\Delta(1232)$ contribution. 
It is also of crucial importance to understand the background contribution, to determine the $N-\Delta$ 
transition form factors in $  \nu_\mu p \to \mu^- p \pi^+ $ and $\bar \nu_\mu n \to \mu^+ n \pi^-$ channels 
which are dominated by $\Delta(1232)$- excitation and its subsequent decay to pions and to interpret
present and future data in (anti)neutrino-nucleon channels 
like $\nu_\mu n \to \mu^- p \pi^0$, $\nu_\mu n \to \mu^- n \pi^+$,
$\bar \nu_\mu p \to \mu^+ n \pi^0$ and $\bar \nu_\mu p \to \mu^+ p \pi^-$. 
A theoretical understanding of presently reported data~\cite{Wilkinson:2014yfa} from reanalysis/reconstruction of 
ANL~\cite{Radecky:1981fn} and BNL~\cite{Kitagaki:1986ct} 
data set in a chiral invariant effective Lagrangian field theoretical
model using $\Delta(1232)$ dominance and non-resonant 
background terms will be highly useful for the purpose of determining various parameters 
like $N-\Delta$ transition form factors, needed to describe 
the basic reaction mechanism of single pion production 
and its application to study nuclear medium and FSI effects. 

In this paper, we have presented the results for the total scattering cross sections
 for pion production from nucleons. The effect of using deuteron target also have been taken into account in a simple model.
We have studied single pion production from free 
nucleons in a model which goes beyond the $\Delta(1232)$ dominance 
model and include the non-resonant contribution from pion pole, nucleon pole and 
contact terms calculated in a chiral invariant field theoretical model. 
The contributions of $I=\frac12$ resonances like $P_{11}$(1440), $D_{13}$(1520), 
$S_{11}$(1535), $S_{11}$(1650) and $P_{13}$(1720)
in second resonance region are also taken into account using a phenomenological Lagrangian. The role of
interference between non-resonant and $\Delta(1232)$ dominant terms is studied. 
The results are compared with the reconstructed data of ANL  and BNL experiments 
reanalyzed by Wilkinson et al.~\cite{Wilkinson:2014yfa} on
$\nu_{\mu}  p \rightarrow \mu^{-} p \pi^{+}$ channel to fix the N-$\Delta$ 
transition form factors and have been applied to study all the pion production channels
from proton and neutron targets which are induced by charged and neutral weak 
currents with neutrino and antineutrino beams.
We have also performed calculations using different cuts on center of mass energy i.e.
($W<1.4 GeV$ and $1.6 GeV$) and compared our results with the available
results of ANL~\cite{Radecky:1981fn} and BNL~\cite{Kitagaki:1986ct} experiments.
For the charged current antineutrino induced $\bar\nu_\mu n \to \mu^+ n \pi^-$
and $\bar\nu_\mu p \to \mu^+ p \pi^-$ processes,
 we have compared our results with the available results of Bolognese et al.~\cite{Bolognese:1979gf}.
For neutral current neutrino induced  process $\nu_{_l} n \to \nu_{_l} p \pi^-$, we have compared the present 
results with the results of Derrick et al.~\cite{Derrick:1980nr}.

In section-\ref{Formalism}, we present the formalism in brief and discuss the non-resonant background mechanism in section-\ref{sec:nr_back} 
 for charged as well as neutral current induced processes. 
While the resonant mechanism are separately discussed in section-\ref{cc1pi} for charged and in section-\ref{nc1pi} for 
neutral current pion production processes.
 The results and their discussions are 
 presented in section-\ref{results}. Finally, we conclude the findings in section-\ref{con}. 
\section{Formalism}\label{Formalism}
The differential scattering cross section for the processes mentioned in Eqs. \ref{eq:CC_all} and \ref{eq:NC_all}, 
 may be written as
\begin{eqnarray}\label{eq:sigma_inelas}
d\sigma &=& \frac{1}{4 \sqrt{(k\cdot p)^2-m_\nu^2M^2}(2\pi)^{5}} \frac{d{\vec k}^{\prime}}{ (2 E_{l})} 
\frac{d{\vec p\,}^{\prime}}{ (2 E^{\prime}_{p})} \frac{d{\vec k}_{\pi}}{ (2 E_{\pi})}
 \delta^{4}(k+p-k^{\prime}-p^{\prime}-k_{\pi})\bar\Sigma\Sigma | \mathcal M |^2,\;\;\;\;\;
\end{eqnarray}
where  $ k( k^\prime) $ is the four momentum of the incoming(outgoing) lepton having energy $E( E^\prime)$ 
while   $p( p^\prime)$ is the four momentum of the incoming(outgoing)
nucleon and the pion momentum is $k_\pi $ having energy $ E_\pi  $. $m_\nu$ is the 
neutrino mass and $M$ is the nucleon mass.
 $ \bar\Sigma\Sigma | \mathcal M |^2  $ is the square of the transition amplitude
 averaged(summed) over the spins of the initial(final) state and may be written as 
\begin{equation}
\label{eq:Gg}
 \mathcal M = \frac{G_F}{\sqrt{2}}\, j_\mu^{(L)} j^{\mu\,{(H)}},
\end{equation}
 where $j_\mu^{(L)}$ and $  j^{\mu\,(H)}$ are the leptonic and hadronic currents, respectively 
 and $G_F$ is the Fermi coupling constant(=$1.166 \times 10^{-5}GeV^{-2}$).
 The weak leptonic current   has $V-A$ structure and is written as
\begin{equation}\label{lep}
j_\mu^{(L)}=\bar u(k^{\prime})\gamma_{\mu}(1 \pm  \gamma_5)u(k),
\end{equation}
where negative sign is for neutrino and positive sign stands for antineutrino 
induced processes.
$  j^{\mu\,(H)}$  describes the hadronic matrix element for  
$W^i + N \rightarrow N^{\prime} + \pi$  interaction     and obtained using an effective 
Lagrangian for $W^i + N \rightarrow N^{\prime} + \pi$ interaction for charged$(W^i \equiv 
W^\pm \; ; i=\pm)$ and neutral current$(W^i \equiv Z^0 \; ; i=0)$ induced processes. 
In this work, we extend our earlier calculations~\cite{Ahmad:2006cy, 
SajjadAthar:2009rc, SajjadAthar:2009rd} which were performed in $\Delta(1232)$ 
dominance model by incorporating non-resonant background terms as well as 
higher resonant terms. The non-resonant background terms involve nucleon and 
pion poles and contact terms calculated using a chiral symmetric Lagrangian for describing their interactions which is obtained in a
non-linear sigma model. The contribution of higher resonances lying in the 
second resonance region beyond the $\Delta(1232)$ resonance are also
included as they may be important in the weak pion production induced by 
(anti)neutrinos of energy $E_{\nu(\bar\nu)}< 2.0 GeV$.
In the following sections, we describe briefly the hadronic matrix element for non-resonant background terms, $\Delta(1232)$ resonance and higher 
resonances.

\begin{figure}[tbh]
\centerline{\includegraphics[height=9cm]{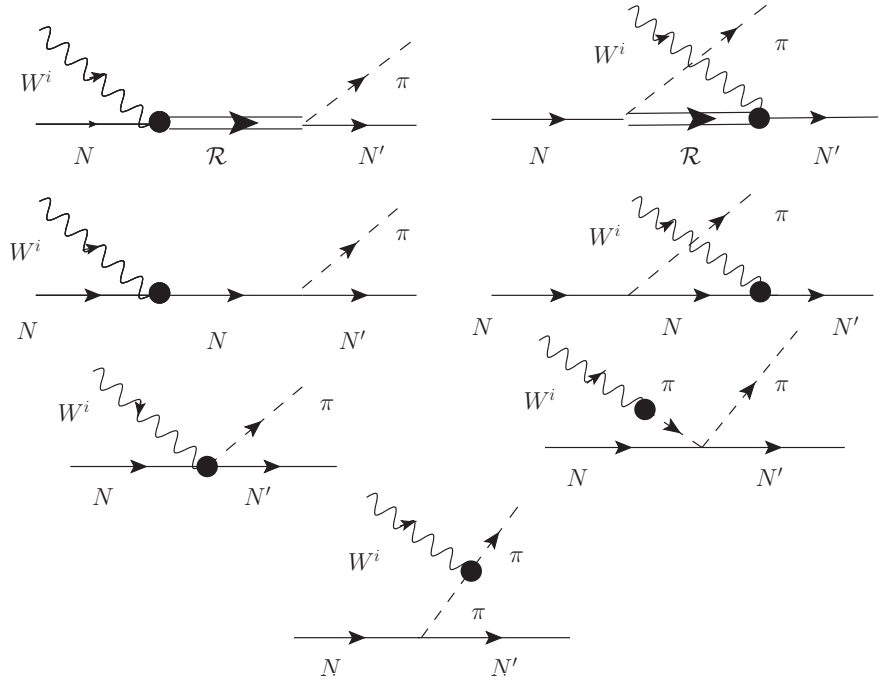}}
\caption{Feynman diagrams contributing to the hadronic current corresponding 
to $W^{i} N \to N^{\prime} \pi^{\pm,0}$, where $(W^i \equiv W^\pm \; ; i=\pm)$ for charged current processes and 
$(W^i \equiv Z^0 \; ; i=0)$ 
for neutral current processes with $N,N^{\prime}=p \;{\rm or}\; n$. First row: direct and cross diagrams for resonance 
production where intermediate term ${\cal R}$ stands for different resonances. Second row: nucleon pole(NP and CNP) 
terms. The contact term(CT) and pion pole(PP) term (third row left to right) and pion in flight(PF)(fourth row)  
contribute to the charged  current processes only and do not contribute to the neutral current processes due to 
their symmetry properties.}
\label{fig:feynmann}
\end{figure}

 The Feynman diagrams which may contribute to the matrix element of the hadronic current are 
shown in Fig.~\ref{fig:feynmann}. The non-resonant background terms include five 
diagrams viz, direct(NP) and cross nucleon pole(CP), 
contact term(CT), pion pole(PP) and pion in flight(PF) terms. For $\Delta(1232)$ resonance 
we have included both direct(s-channel) and cross(u-channel) diagrams. 
Apart from $\Delta(1232)$ resonance, we have also taken contributions from 
$P_{11}$(1440), $S_{11}$(1535) and $S_{11}$(1650)
spin half resonances and $D_{13}$(1520) and $P_{13}$(1720) spin three-half 
resonances and considered both s-channel and u-channel contributions. 

In the following sections, we present the formalism in brief which has been used for the 
non-resonant background terms and the resonant spin half and spin three-half contributions to the 
one pion production processes. 

\subsection{Non-resonant background contribution}\label{sec:nr_back}
The contribution from the non-resonant background terms in the case of 
charged$(W^i \equiv W^\pm \; ; i=\pm)$ and neutral$(W^i \equiv Z^0 \; ; i=0)$ 
current reaction $W^i N \to N^{\prime}\pi$ may be obtained using non-linear sigma model 
based on the works of Hernandez et al.~\cite{Hernandez:2007qq}. 
In lowest order, the contributions to the hadronic current  are written in a model independent way as
\begin{eqnarray} \label{eq:background}
j^\mu\big|_{NP} &=& 
a~\mathcal{A}^{NP}
  \bar u(\vec{p}\,') 
 \slashchar{k}_\pi\gamma_5\frac{\slashchar{p}+\slashchar{q}+M}{(p+q)^2-M^2+ i\epsilon}\left [V^\mu_N(q)-A^\mu_N(q) \right]  
u(\vec{p}\,),\nonumber \\\nonumber \\
j^\mu\big|_{CP} &=& 
a~\mathcal{A}^{CP}
  \bar u(\vec{p}\,') \left [V^\mu_N(q)-A^\mu_N(q) \right]
\frac{\slashchar{p}'-\slashchar{q}+M}{(p'-q)^2-M^2+ i\epsilon} \slashchar{k}_\pi\gamma_5  u(\vec{p}\,),\nonumber \\\nonumber \\
j^\mu\big|_{CT} &=&
a~\mathcal{A}^{CT}
  \bar u(\vec{p}\,') \gamma^\mu\left (
  g_A  f_{CT}^V(Q^2)\gamma_5 - f_\rho\left((q-k_\pi)^2\right) \right ) u(\vec{p}\,),\nonumber \\\nonumber \\
j^\mu\big|_{PP} &=& 
a~\mathcal{A}^{PP}f_\rho\left((q-k_\pi)^2\right)
  \frac{q^\mu}{m_\pi^2+Q^2}
  \bar u(\vec{p}\,')\ \slashchar{q} \ u(\vec{p}\,),\nonumber \\ \nonumber \\
j^\mu\big|_{PF} &=& 
a~\mathcal{A}^{PF}f_{PF}(Q^2)
  \frac{(2k_\pi-q)^\mu}{(k_\pi-q)^2-m_\pi^2}
  2M\bar u(\vec{p}\,')  \gamma_5 u(\vec{p}\,),\label{eq:eqscc}
\end{eqnarray}
with $ a = \cos \theta_C$ for charged current process and $a$ = 1 for neutral current process.
$q$ is the four momentum transfer(=$k-k^\prime$), $q^2(=-Q^2) \le 0$ and $k_\pi$ is the pion momentum.
M is the mass of nucleon and $m_\pi$   is the mass of pion.
The constant factor $\mathcal{A}^i, \; i=NP,CP, CT, PP \; {\rm and} \; PF$, and  
are tabulated in Table--\ref{tab:born_para}. 

The  vector$(V^\mu_N(q))$ 
and axial vector$(A^\mu_N(q))$ currents for nucleon pole diagrams 
in the case of charged and neutral current interactions are
 calculated neglecting second class currents and are given by,
\begin{eqnarray}\label{eq:vec_curr}
V^\mu_N(q)&=&\tilde f_1(Q^2)\gamma^\mu+\tilde f_2(Q^2)i\sigma^{\mu\nu}\frac{q_\nu}{2M} \\
\label{eq:axi_curr}
A^\mu_N(q)&=& \left(\tilde f_A(Q^2)\gamma^\mu 
+ \tilde f_P(Q^2) \frac{q^\mu}{M}  \right)\gamma^5,
\end{eqnarray}
where  $\tilde f_{1,2}(Q^2)$ and $\tilde f_{A,P}(Q^2)$ are the vector and axial vector form factors for  nucleons. 
 In the case of charged current process, the form factors $\tilde f_{1,2}(Q^2)$ 
are expressed in terms of isovector($f_{1,2}^V(Q^2)$) form factors as:
\begin{equation}\label{f1v_f2v}
\tilde f_{1,2}(Q^2) \longrightarrow f_{1,2}^V(Q^2)=f_{1,2}^p(Q^2)- f_{1,2}^n(Q^2), 
\end{equation}
 where $f_{i}^{p,n}(Q^2); \;\; i=1,2$ are the Dirac$(i=1)$ and Pauli$(i=2)$ form factors of nucleons. These form factors are in turn 
 expressed in terms of the experimentally 
determined Sach's electric $G_E^{p,n}(Q^2)$ and magnetic $G_M^{p,n}(Q^2)$ form factors~\cite{Galster:1971kv}.   
While in  the case of neutral current process, the form factors are expressed as:
\begin{align}
\tilde f_{1,2}(Q^2) &  \mathrel{\mathop{\longrightarrow}^{\mathrm{for \; p}}}  \tilde f_{1,2}^{p}(Q^2)=\left(\frac12 - 2 \sin^2 \theta_W\right)
f_{1,2}^{p}(Q^2)- \frac12 f_{1,2}^{n}(Q^2) \nonumber \\ 
\tilde f_{1,2}(Q^2) &\mathrel{\mathop{\longrightarrow}^{\mathrm{for \; n}}} \tilde f_{1,2}^{n}(Q^2)=\left(\frac12 - 2 \sin^2 \theta_W\right)
f_{1,2}^{n}(Q^2)- \frac12 f_{1,2}^{p}(Q^2).
\end{align}
where $\theta_W$ is the Weinberg angle.
On the other hand, the axial form factor($\tilde f_{A}(Q^2)$)  is generally taken to be of dipole form and is given by
\begin{equation}\label{fa}
\tilde f_{A}(Q^2)=f_A(Q^2)=f_A(0)~\left[1+\frac{Q^2}{M_A^2}\right]^{-2},
\end{equation}
for charged current and by 
\begin{equation}\label{fa_nc}
\tilde f_{A}(Q^2)=\tilde f^{p,n}_A(0)~\left[1+\frac{Q^2}{M_A^2}\right]^{-2},
\end{equation}
for neutral current.
Here $f_A(0)$ is the axial charge and is  obtained from the quasielastic neutrino and antineutrino scattering
as well as from the pion electro-production data. We have used $f_A(0)$=--1.267 and
the axial dipole mass $M_A$=1.026GeV, which is the world average value~\cite{Bernard:2001rs}, in the numerical calculations. 
 For the neutral current induced reaction, $\tilde f^{p,n}_{A}(Q^2)= \pm \frac12 f_{A}(Q^2)$, where the plus(minus) sign stands for proton(neutron) target. 

The next contribution from the axial part comes from the pseudoscalar form factor $ \tilde f_P(Q^2)(= f_P(Q^2))$, 
the determination of which is based on PCAC and pion pole dominance 
and is related to $f_A(Q^2)$ through the relation 
\begin{equation}\label{fp}
f_P (Q^2)=\frac{2M^2 \; f_A (Q^2)}{m_\pi^2+Q^2}.
\end{equation}
 The contribution of this form being proportional to lepton mass vanishes for the neutral current processes.
 The hadronic current for NC processes have the contribution only from nucleon 
pole terms(s and u channels), while CC processes have contribution 
from all the diagrams viz. NP, CP, CT, PP and PF terms.

 In order to conserve vector current for CC processes at the weak vertex, 
the  two form factors viz. $f_{PF}(Q^2)$ and 
$f_{CT}^{V}(Q^2)$ are expressed in terms of the isovector nucleon form factor as~\cite{Hernandez:2007qq}
\begin{equation}
 f_{PF}(Q^2) = f_{CT}^{V}(Q^2) = 2 f_{1}^V(Q^2).
\end{equation}
The $\pi \pi NN$ vertex has the dominant $\rho$--meson cloud contribution and following Ref.~\cite{Hernandez:2007qq}, 
we have introduced $\rho-$form factor $(f_{\rho}(Q^2))$ at $\pi \pi NN$ vertex and taken it to be of monopole form:
\begin{equation}
 f_{\rho}(Q^2) = \frac{1}{1+Q^2/m_{\rho}^2}; \qquad \qquad {\rm with } \; m_\rho = 0.776 GeV.
\end{equation}
$f_{\rho}(Q^2)$ also  has been used with axial part of the CT diagram in order to be consistent with 
the assumption of PCAC. 

In the next section, we will discuss the formalism for charged and neutral current 
neutrino(antineutrino) induced processes for pion production through resonance excitations.

\subsection{Charged current neutrino(antineutrino) induced processes}\label{cc1pi}

\begin{table*}[h]
  \begin{center}
  \vspace{1cm}
    \renewcommand{\arraystretch}{2.}
    \begin{tabular*}{147mm}{@{\extracolsep{\fill}}|c|ccc|ccc|cccc|}
      \noalign{\vspace{-8pt}}
      \hline \hline
      Constant term $\rightarrow$         & \multicolumn{3}{c|}{$\mathcal{A}$(CC $\nu$)}&\multicolumn{3}{c|}{$\mathcal{A}$(CC $\bar \nu$)}
      &\multicolumn{4}{c|}{$\mathcal{A}$(NC $\nu(\bar\nu)$)}     \\ \hline
      Final states  $\rightarrow$  &   $p\pi^{+}$ & $n\pi^{+}$ & $p\pi^{0}$&
      $n\pi^{-}$ & $n\pi^{0}$ & $p\pi^{-}$&
      $n\pi^{+}$ & $p\pi^{0}$ & $p\pi^{-}$&$n\pi^{0}$ \\ \hline
      NP  &   0   & $\frac{-ig_A}{\sqrt{2}f_\pi}$& $\frac{-ig_A}{f_\pi}$ &
              0   & $\frac{ig_A}{\sqrt{2}f_\pi}$& $\frac{-ig_A}{f_\pi}$ &
              $\frac{-ig_A}{\sqrt{2}f_\pi}$& $\frac{-ig_A}{f_\pi}$ & $\frac{-ig_A}{\sqrt{2}f_\pi}$& $\frac{ig_A}{f_\pi}$\\
      CP &   $\frac{-ig_A}{f_\pi}$  &0 & $\frac{ig_A}{\sqrt{2}f_\pi}$  &
              $\frac{-ig_A}{f_\pi}$& $\frac{-ig_A}{\sqrt{2}f_\pi}$ &  0 &
              $\frac{-ig_A}{\sqrt{2}f_\pi}$& $\frac{-ig_A}{f_\pi}$ & $\frac{-ig_A}{\sqrt{2}f_\pi}$& $\frac{ig_A}{f_\pi}$\\
      CT  &   $\frac{-i}{\sqrt{2}f_\pi}$ &$\frac{i}{\sqrt{2}f_\pi}$ & $\frac{i}{2 f_\pi}$   &
              $\frac{-i}{\sqrt{2}f_\pi}$ & $\frac{i}{f_\pi}$ & $\frac{i}{\sqrt{2}f_\pi}$&
              - & - & - & -\\
      PP  &   $\frac{i}{\sqrt{2}f_\pi}$ & $\frac{-i}{\sqrt{2}f_\pi}$& $\frac{-i}{2 f_\pi}$    &
              $\frac{i}{\sqrt{2}f_\pi}$&$\frac{i}{f_\pi}$& $\frac{-i}{\sqrt{2}f_\pi}$  &
              - & - & - & -\\
      PF  &   $\frac{-i}{\sqrt{2}f_\pi}$ &$\frac{i}{\sqrt{2}f_\pi}$ & $\frac{i}{2 f_\pi}$   &
              $\frac{-i}{\sqrt{2}f_\pi}$&$\frac{-i}{f_\pi}$&$\frac{-i}{\sqrt{2}f_\pi}$ &
              - & - & - & -\\
      \hline \hline
    \end{tabular*}
  \end{center}
    \caption{The values of constant term($\mathcal{A}^i$) appearing in Eq.~\ref{eq:eqscc}, where $i$ corresponds to 
    the nucleon pole(NP), cross nucleon pole(CP), contact term(CT), pion pole(PP) and pion in flight(PF) terms. $f_\pi$ is pion weak decay constant and $g_A$ is 
    axial nucleon coupling.}
    \label{tab:born_para}
\end{table*}
 Furthermore, we have also studied the contribution of other higher resonances to the single pion production
channel which may also contribute along with the dominant $\Delta(1232)$ resonance, formalism for which has been discussed in the next section.
\subsubsection{Resonant Contribution}\label{sec3}
Besides the non-resonant background contribution to the pion production there are several resonances which 
may contribute along with the dominant $\Delta(1232)$ resonance channel. 
The basic neutrino(antineutrino) induced reactions for pion production through resonance excitations are the following:
\begin{eqnarray}\label{eq:nucc_res}
\nu_{_l}(k) + N(p) \rightarrow l^{-}(k^{\prime}) + {\cal R}(p_R) & \nonumber \\
 {\rotatebox[origin=c]{180}{{\huge$\Lsh$}}} \quad \; &
\hspace*{-0.65cm} N^\prime (p^\prime) + \pi (k_\pi) 
\end{eqnarray}
\begin{eqnarray}\label{eq:anucc_res}
\bar\nu_{_l}(k) + N(p) \rightarrow l^{+}(k^{\prime}) + {\cal R}(p_R) & \nonumber \\
 {\rotatebox[origin=c]{180}{{\huge$\Lsh$}}} \quad \; &
\hspace*{-0.65cm} N^\prime (p^\prime) + \pi (k_\pi) 
\end{eqnarray}
where ${\cal R}$ stands for $\Delta(1232)$ and/or other higher resonances$(R)$ which contribute to the pion production. 
In the next section, we briefly describe our model to include different 
resonances for the charged current and the neutral current induced reactions. 
In the present work, we have included six resonances, properties of
which are summarized in Table-\ref{tab:included_resonances}. It may be noticed from the table that considered resonances
are spin $\frac32$  and spin $\frac12$ resonant states with positive or negative parity. 
We shall discuss in brief the structure of the current for these two resonant states.\\

{\bf A. Spin $\frac{3}{2}$ resonances}

 The general structure for the hadronic current for spin three-half resonance excitation 
   is determined by the following equation~\cite{LlewellynSmith:1971zm} 
\begin{eqnarray}\label{eq:had_current_3/2}
J_{\mu}^{\frac{3}{2}}=\bar{\psi}^{\nu}(p') \Gamma_{\nu \mu}^{\frac32} u(p), 
\end{eqnarray}
where u(p) is the Dirac spinor for nucleon, ${\psi}^{\mu}(p)$ is the Rarita-Schwinger spinor for spin three-half particle and 
$\Gamma_{\nu \mu}^{\frac32}$ has the following general structure for the positive and negative parity states : 
\begin{eqnarray}\label{eq:vec_3half_pos}
  \Gamma_{\nu \mu }^{\frac{3}{2}^+} &=& \left[{V}_{\nu \mu }^\frac{3}{2} - {A}_{\nu \mu }^\frac{3}{2}\right]  \gamma_5  \nonumber\\
  \Gamma_{\nu \mu }^{\frac{3}{2}^-} &=&  {V}_{\nu \mu }^\frac{3}{2} - {A}_{\nu \mu }^\frac{3}{2} ,
\end{eqnarray}
where $V_{\frac32}(A_{\frac32})$ is the vector(axial-vector) current for spin three-half resonances.  
The vector and the axial-vector part of the currents are given by
\begin{eqnarray}\label{eq:vec_axial}
  V_{\nu \mu }^{\frac{3}{2}} &=& \left[ \frac{{\tilde C}_3^V}{M} (g_{\mu \nu} \slashchar{q} \, - q_{\nu} \gamma_{\mu})+
  \frac{{\tilde C}_4^V}{M^2} (g_{\mu \nu} q\cdot p' - q_{\nu} p'_{\mu}) 
  + \frac{{\tilde C}_5^V}{M^2} (g_{\mu \nu} q\cdot p - q_{\nu} p_{\mu}) + 
  g_{\mu \nu} {\tilde C}_6^V\right]    \nonumber\\
  A_{\nu \mu }^{\frac{3}{2}} &=&- \left[ \frac{{\tilde C}_3^A}{M} (g_{\mu \nu} \slashchar{q} \, - q_{\nu} \gamma_{\mu})+
  \frac{{\tilde C}_4^A}{M^2} (g_{\mu \nu} q\cdot p' - q_{\nu} p'_{\mu})+
 {{\tilde C}_5^A} g_{\mu \nu}
  + \frac{{\tilde C}_6^A}{M^2} q_{\nu} q_{\mu}\right] \gamma_5
\end{eqnarray}
where ${\tilde C}^V_i$ and ${\tilde C}^A_i$ are the vector and axial charged current transition form factors which are functions of $Q^2$.

From the conserved vector current hypothesis one takes ${\tilde C}_6^V(Q^2)=0$.
First, we shall discuss the expression of the form factors for the $\Delta(1232)$
resonance and write ${\tilde C}^V_i=C^V_i$ and ${\tilde C}^A_i=C^A_i$. Now for the $\Delta(1232)$ resonance, the other three vector form
factors $C_i^V,i=3,4,5$ are given in terms of the isovector electromagnetic form
factors for $p \rightarrow \Delta^+$  transition and the parameterization of which are taken from the Ref.~\cite{Lalakulich:2006sw}, 
\begin{eqnarray}\label{vecff}
C_3^V(Q^2) &=& \frac{2.13}{(1+Q^2/M_V^2)^2}\times
\frac{1}{1+\frac{Q^2}{4M_V^2}},\nonumber \\
\qquad C_4^V(Q^2) &=& \frac{-1.51}{(1+Q^2/M_V^2)^2}\times \frac{1}{1+\frac{Q^2}{4M_V^2}},\nonumber \\
\qquad C_5^V(Q^2) &=& \frac{0.48}{(1+Q^2/M_V^2)^2}\times
\frac{1}{1+\frac{Q^2}{0.776M_V^2}} \label{eq:c3v}
\end{eqnarray}
with the vector dipole mass taken as $M_V= 0.84$ GeV. 
 
The axial vector form factors $C_i^A(Q^2), \; (i=3,4,5)$ are generally determined by using the hypothesis of PCAC 
with pion pole dominance through the off diagonal Goldberger-Trieman relation or obtained in
quark model calculations~\cite{Liu:1995bu,Hemmert:1994ky}. The early analysis of weak pion production 
experiments at ANL~\cite{Radecky:1981fn} and BNL~\cite{Kitagaki:1986ct}
used Adler's model~\cite{Adler:1968tw} as developed by Schreiner and von Hippel~\cite{Schreiner:1973mj} to determine these form factors which are consistent with the 
hypothesis of PCAC and generalized Goldberger-Trieman relation.
These considerations give $C_6^A(Q^2)$ in terms of  $C_5^A(Q^2)$ and $C_5^A(0)$ in terms of $f_{\Delta N \pi}$ as: 
\begin{align}\label{c6-c5}
 C_6^A(Q^2) =& C_5^A(Q^2) \frac{M^2}{Q^2 + m_\pi^2} \\
 C_5^A(0) =& f_\pi \frac{ f_{\Delta N \pi}  }{2 \sqrt3 M },
\end{align}
where $f_{\Delta N \pi}$ is the $\Delta N \pi$ coupling strength  for $\Delta \rightarrow N \pi$ decay.

The $Q^2$ dependence of $C_3^A(Q^2)$ and $C_4^A(Q^2)$ are obtained in Adler's model as\cite{Adler:1968tw,Schreiner:1973mj} 
\begin{equation}\label{c4-c3}
 C_4^A(Q^2) = -\frac{1}{4}C_5^A(Q^2);  \qquad            C_3^A(Q^2) =0.
\end{equation}

The $Q^2$ dependence of $C_5^A$ is parameterized by Schreiner and von Hippel~\cite{Schreiner:1973mj} in the Adler's model~\cite{Adler:1968tw} 
 and is given by 
 \begin{eqnarray}\label{c5aq}
  C_5^A(Q^2) &=&  \frac{C_5^A(0) \left( 1+ \frac{a\, Q^2}{b~+~Q^2} \right)}{\left( 1 + Q^2 /M_{A\Delta}^2 \right)^2} 
\end{eqnarray}
with $a$ and $b$ are determined from the experiments and found to be $a=-1.21$ and $b=2$~\cite{Radecky:1981fn,Kitagaki:1990vs}.
 $M_{A\Delta}$ is the axial dipole mass.

 The axial vector form factors as discussed in Eqs.\ref{c6-c5} and \ref{c5aq} along with the vector form factors given in Eq.\ref{vecff}, have been used to analyze
 the present experimental cross  sections for weak pion production. 
 Most of the recent theoretical calculations~\cite{Lalakulich:2006sw,Hernandez:2007qq,Leitner:2006ww,Lalakulich:2010ss} use a simpler modification to the dipole form viz.
\begin{equation}
C_5^A(Q^2) = \frac{C_5^A(0)}{ \left( 1 + Q^2 /M_{A\Delta}^2 \right)^2 } \; \frac{1}{ 1 + Q^2/(3 M_{A\Delta}^2)}.
\end{equation}
With the non-vanishing axial vector form factors determined in terms of $C_5^A(Q^2)$ and the vector 
form factors determined from electron scattering experiments, the weak pion production cross section is described  
in terms of $C_5^A(Q^2)$ with the parameters $C_5^A(0)$ and axial mass $M_{A\Delta}$. We chose $M_{A\Delta}=1.026 {\rm GeV}$ corresponding 
to the world average 
 value obtained from the experimental analysis of quasielastic scattering events~\cite{Bernard:2001rs}, and then vary $C_5^A(0)$ to obtain a good 
 description of reanalyzed data~\cite{Wilkinson:2014yfa} of ANL and BNL experiments for $\nu_\mu  p \rightarrow \mu^-   p   \pi^+$ reaction.
 While fitting the reanalyzed data for the reaction $\nu_\mu  p \rightarrow \mu^- p \pi^+$, the contributions to the cross section 
 is predominantly obtained from $\Delta(1232)$ resonant terms and the background terms have a little contribution. This has been further 
 discussed in Section~\ref{results}.

One may write the most general form of the hadronic current for the s-channel(direct diagram) and u-channel(cross diagram) processes
where a resonant state $R^{\frac32}$ is produced and decays to a pion in the final state as
\begin{eqnarray}\label{eq:res_had_current}
j^\mu\big|_{R}^{\frac32} &=& i \; a\; \mathcal{C^{R}} 
   \frac{k_\pi^\alpha}{p_R^2-M_R^2+ i M_R \Gamma_R}
   \bar u(\vec{p}\,') P_{\alpha\beta}^{3/2}(p_R) \Gamma^{\beta\mu}_{\frac32}(p,q)
   u(\vec{p}\,),\quad p_R=p+q, \nonumber  \\ 
 j^\mu\big|_{C R}^{\frac32} &=& i \; a\; \mathcal{C^{R}} 
   \frac{k_\pi^\beta}{p_R^2-M_R^2+ i M_R \Gamma_R}
   \bar u(\vec{p}\,')  {\hat \Gamma}^{\mu\alpha}_{\frac32}(p',-q) P_{\alpha\beta}^{3/2}(p_R) 
   u(\vec{p}\,),\quad p_R=p'-q, \nonumber \\
\end{eqnarray}
where $a=\cos \theta_c$ for the charged current process and $a= 1$ for the neutral 
current process and $ \mathcal{C^{R}} $ is the coupling strength for ${\cal R} \to N\pi$,
${\cal R}\equiv \Delta$ or any other resonance $R$,
determined from partial decay widths. $M_R$ is the mass of resonance and $\Gamma_R$ is the resonant decay width.
 These resonances are generally off-shell and their off-shell effects are also taken into account. 
$P_{\alpha\beta}^{3/2}$ is spin three-half projection operator and is given by
\begin{equation}
P_{\alpha\beta}^{3/2}=- \left(\slashchar{p^\prime} \, + M_R \right) \left( g_{\alpha \beta}
- \frac{2}{3} \frac{p'_{\alpha } p'_{\beta}}{M_R^2} 
+ \frac{1}{3} \frac{p'_{\alpha } \gamma_{\beta} - p'_{\beta } \gamma_{\alpha}}{M_R} 
- \frac{1}{3} \gamma_{\alpha} \gamma_{\beta} \right).
\end{equation}

\begin{figure}
\includegraphics[width=14cm,height=4.5cm]{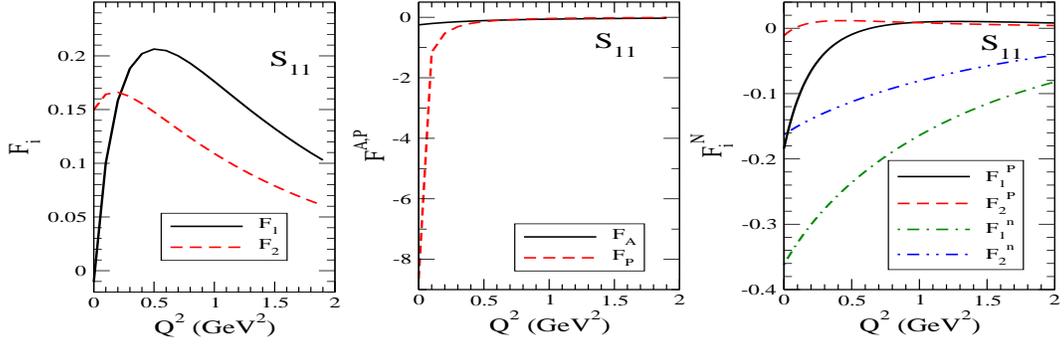}
\caption{$Q^2$ dependence of different form factors of $S_{11}$ resonance. In the left panel the form of isovector form factors 
$F_1^V$ and $F_2^V$ are shown the form of which are given in Eq.~\ref{eq:f12vec_res}.
In the middle panel we have shown the axial form factors $F_{A,P}$ where for $F_{A}$ we took the dipole form 
and the $F_P$ is obtained from Eq.~\ref{eq:fp_res_spinhalf}. 
In the right panel we have shown the explicit dependence of $F_i^{p,n}, \; i=1,2$ which may be obtained from the 
helicity amplitudes as given in Eq.~\ref{eq:hel_spin_12}.}
\label{fig:S11_FF}
\end{figure}

Apart from $\Delta(1232)$ resonance, we have also included other higher spin $\frac32$ resonances like  $D_{13}(1520)$ and $P_{13}(1720)$.
 The structure of the matrix element for the hadronic current is given in Eq.~\ref{eq:res_had_current}
 and the weak vertex for positive and negative parity states are 
given in Eqs.~\ref{eq:vec_3half_pos}.  
 The vector and axial vector pieces  are written in analogy with $\Delta(1232)$ resonance which are given in Eq.~\ref{eq:vec_axial}
 with corresponding form factors, $\tilde{C}^V_i$ and $\tilde{C}^A_i$, defined for each resonances. 

The isovector $\tilde{C}_i^V , i=3,4,5,6$ form factors for $D_{13}(1520)$ and $P_{13}(1720)$, which have $J=\frac32 , \, I=\frac12  $, 
are written in terms of the charged and neutral form factors ($C^{p,n}_i(Q^2)$) through a simple relation~\cite{Leitner:2006ww},
\begin{equation}\label{eq:civ_NC}
{\tilde C}_i^V = C^{p}_i  - C^{n}_i ; \;\; \qquad i = 3,4,5,6 \, . 
\end{equation}
The vector form factors $C^{p,n}_i(Q^2)$ are related with the helicity amplitudes for which
the $Q^2$ dependence is parameterized as~\cite{Tiator:2011pw}  
\begin{equation}\label{eq:ffpar}
{\mathcal A}_{\alpha}(Q^2) = {\mathcal A}_{\alpha}(0) (1+a_1 Q^2+a_2 Q^4
+a_3 Q^6+a_4 Q^8)\, e^{-b_1 Q^2} 
\end{equation}
where $ {\mathcal A}_{\alpha}(Q^2)$ is the helicity amplitude; $A_{\frac32(\frac12)}(Q^2) \; {\rm and/or}\; S_{\frac12}(Q^2)$ and 
parameters $ {\mathcal A}_{\alpha}(0)$ are generally determined
by a fit to the   photoproduction  data of the corresponding resonance. 
While the parameters $a_i \;(i=1-4)$ and
$b_1$ for each amplitude  are obtained from electroproduction data available at different  $Q^2$. 

The relations between the vector form factors $C^{p,n}_i(Q^2)$ and helicity amplitudes are given as~\cite{Drechsel:2007if}:

\begin{eqnarray}
A_{\frac{3}{2}}^{p,n} &=& \sqrt{\frac{\pi \alpha}{M}  \frac{(M_R \mp M)^2+Q^2}{M_R^2-M^2}}  
\left[ \frac{C^{p,n}_3}{M} (M \pm  M_R) \pm  \frac{C^{p,n}_4}{M^2} \frac{M_R^2-M^2 - Q^2}{2} \right.\nonumber\\
&\pm& \left.
\frac{C^{p,n}_5}{M^2} \frac{M_R^2-M^2 + Q^2}{2} \right] \nonumber \\
 A_{\frac{1}{2}}^{p,n} &=&\sqrt{\frac{\pi \alpha}{3 M}  \frac{(M_R \mp M)^2+Q^2}{M_R^2-M^2}}  
 \left[ \frac{C^{p,n}_3}{M} \frac{M^2+M M_R +Q^2}{M_R} -  \frac{C^{p,n}_4}{M^2} \frac{M_R^2-M^2 -
Q^2}{2} \right. \nonumber \\
&-& \left.  \frac{C^{p,n}_5}{M^2} \frac{M_R^2-M^2 + Q^2}{2} \right] \nonumber \\ 
S_{\frac{1}{2}}^{p,n} &=& \pm \sqrt{\frac{\pi \alpha}{6 M}  \frac{(M_R \mp  M)^2+Q^2}{M_R^2-M^2}} 
\frac{\sqrt{ Q^4 + 2 Q^2 (M_R^2 + M^2) + (M_R^2-M^2)^2    }}{M_R^2} \nonumber \\
 &\times&  \left[ \frac{C^{p,n}_3}{M} M_R +  \frac{C^{p,n}_4}{M^2} M_R^2 +  \frac{C^{p,n}_5}{M^2}
\frac{M_R^2+M^2 + Q^2}{2}  \right],
\end{eqnarray}
where  $A_{\frac32, \frac12}(Q^2)$ and $S_{\frac12}(Q^2)$ are the amplitudes 
corresponding to the transverse and longitudinal polarizations, respectively 
and are parameterized at different $Q^2$ using Eq.~\ref{eq:ffpar}. 
Once the parameters $a_i$ and $b_1$ are fixed (Tables~\ref{tab:param-p1}, \ref{tab:param-p2}, \ref{tab:param-n})  
for $A_{\frac32, \frac12}(Q^2)$ and $S_{\frac12}(Q^2)$ amplitudes, one gets the form factors $C^{p,n}_i(Q^2)$.
 
The form factors ${\tilde C}_i^A(Q^2), \; (i=3,4,5,6)$ corresponding to the axial current have 
 not been studied in the case of higher resonances. The earlier calculations have used PCAC to determine
 ${\tilde C}_5^A(Q^2)$ and ${\tilde C}_6^A(Q^2)$ and taken 
 other form factors to be zero. In view of this, we have also taken a simple model 
for the determination of the axial form factors 
based on PCAC and Goldberger-Trieman relation and use the relation between ${\tilde C}_5^A(Q^2)$ and ${\tilde C}_6^A(Q^2)$ given in Eq.~\ref{c6-c5} 
to write ${\tilde C}_6^A(Q^2)$ in terms of ${\tilde C}_5^A(Q^2)$. 
 
 For ${\tilde C}_5^A(Q^2)$ a dipole form has been assumed 
 \begin{equation}\label{c5a-r}
{\tilde C}_5^A(Q^2) = \frac{{\tilde C}_5^A(0)}{ \left( 1 + Q^2 /{M_A^{\it R}}^2 \right)^2 } 
\end{equation}
 with ${\tilde C}_5^A(0)= -2 f_\pi \frac{f_{R N \pi}}{m_\pi}$, $f_{R N \pi}$ is the coupling for
 $R \rightarrow N \pi$ decay for each resonance $R$. 
  $M_{A}^{\it R}$ is taken as $1.026 {\rm GeV}$. The value of ${\tilde C}_5^A(0)$ for each 
  resonance has been given in Table-2. ${\tilde C}_3^A(Q^2)$ as well as ${\tilde C}_4^A(Q^2)$ are taken as zero.

In the next section, we briefly discuss the inputs of hadronic current for spin $\frac12$ resonance. \\

{\bf {B. Spin $\frac12$ resonances}}\label{spin12reso}\\

The hadronic current for the spin $\frac12$ resonant state is given by  
\begin{eqnarray}\label{had_curr_1/2}
j^{\mu}_{\frac{1}{2}}=\bar{u}(p') \Gamma^\mu_{\frac12} u(p), 
\end{eqnarray}
where $u(p)$ and $\bar u(p^\prime)$ are respectively the Dirac spinor and adjoint Dirac spinor for spin $\frac{1}{2}$ particle and 
$\Gamma^\mu_\frac12$ is the vertex function which for a positive parity state is given by 
\begin{align}\label{eq:vec_half_pos}
  \Gamma^{\mu}_{\frac{1}{2}^+} &= {V}^{\mu}_\frac{1}{2} -  {A}^{\mu}_\frac{1}{2}  
  \end{align}
and for a negative parity is given by  
\begin{align}\label{eq:vec_half_neg}
  \Gamma^{\mu}_{\frac{1}{2}^-} &= \left[ {V}^{\mu}_\frac{1}{2} -  {A}^{\mu}_\frac{1}{2}\right] \gamma_5  
  \end{align}
where $V^{\mu}_{\frac{1}{2}}$ represents the vector current and $A^{\mu}_{\frac{1}{2}}$ 
 represents the axial vector current.
 
These currents are parameterized in terms of vector$(F_i(Q^2) (i=1,2))$ and axial 
vector$({F_A}(Q^2)$ and ${F_P}(Q^2))$ form factors and are written as, 
\begin{align}  \label{eq:vectorspinhalfcurrent}
  V^{\mu}_{\frac{1}{2}} & =\left[\frac{{F_1}(Q^2)}{(2 M)^2}
  \left( Q^2 \gamma^\mu + \slashchar{q} q^\mu \right) + \frac{F_2(Q^2)}{2 M} 
  i \sigma^{\mu\alpha} q_\alpha \right] \gamma_5  \\
    \label{eq:axialspinhalfcurrent}
  A^{\mu}_\frac{1}{2} &= - {F_A}(Q^2) \gamma^\mu  - 
  \frac{F_P(Q^2)}{M} q^\mu,
\end{align}
 where $F_i(Q^2)$ ($i=1,2$) are the isovector form factors which in turn are expressed in terms of Dirac and Pauli form factors for spin $\frac{1}{2}$ resonances 
  for charged $(F_{1,2}^{p})$ and neutral $(F_{1,2}^n)$ states:
\begin{equation}\label{eq:f12vec_res}
 F_i(Q^2) = F_i^p(Q^2) - F_i^n(Q^2), \quad \quad i=1,2
\end{equation}
The form factors $F^{p,n}_{i}(Q^2)$ are derived from helicity amplitudes extracted from
real and/or virtual photon scattering experiments.

The explicit relations between the form factors
$F_i^{p,n}(Q^2)$ and the helicity amplitudes $A_{\frac{1}{2}}^{p,n}(Q^2)$ and $S_{\frac{1}{2}}^{p,n}(Q^2)$
are given by\cite{Leitner:2008ue}
\begin{eqnarray}\label{eq:hel_spin_12}
A_\frac{1}{2}^{p,n}&=& \sqrt{\frac{2 \pi \alpha}{M} \frac{(M_R \mp M)^2+Q^2}{M_R^2 - M^2}} \left[  \frac{Q^2}{4 M^2}
F_1^{p,n} + \frac{M_R \pm M}{2 M} F_2^{p,n} \right] \nonumber \\
S_\frac{1}{2}^{p,n}&=&\mp~\sqrt{\frac{ \pi \alpha}{M} \frac{(M \pm M_R)^2+Q^2}{M_R^2 - M^2}}
 \frac{(M_R \mp M)^2 +Q^2}{4 M_R M} \left[
\frac{M_R \pm M}{2 M} F_1^{p,n} - F_2^{p,n}\right],
\end{eqnarray}
where the upper sign represent the positive parity state and the lower sign denotes the negative parity state. 
$M_R$ is the mass of corresponding resonance  and
$F^{p,n}_{1,2}(Q^2)$ are electromagnetic transition form factors.
The $Q^2$ dependence of the helicity amplitudes is given by Eq.~\ref{eq:ffpar}.

The axial current consists of two from factors viz. $F_A(Q^2)$ and $F_P(Q^2)$. 
The form factor $F_A(Q^2)$ and $F_P(Q^2)$ are determined assuming the 
 the hypothesis of PCAC and pion pole dominance through the off diagonal Goldberger-Trieman relation for N $\rightarrow$ {\it R} transition, which gives
\begin{equation}
F_A(0)= -2 f_\pi \frac{f_{R\frac12}}{m_\pi},
\end{equation}
 where $f_{R\frac12}$ is the coupling strength for $R_{\frac12} \to N \pi$ decay and $F_A(0)$ is the axial charge.
 
\begin{figure}
\includegraphics[width=14cm,height=4.5cm]{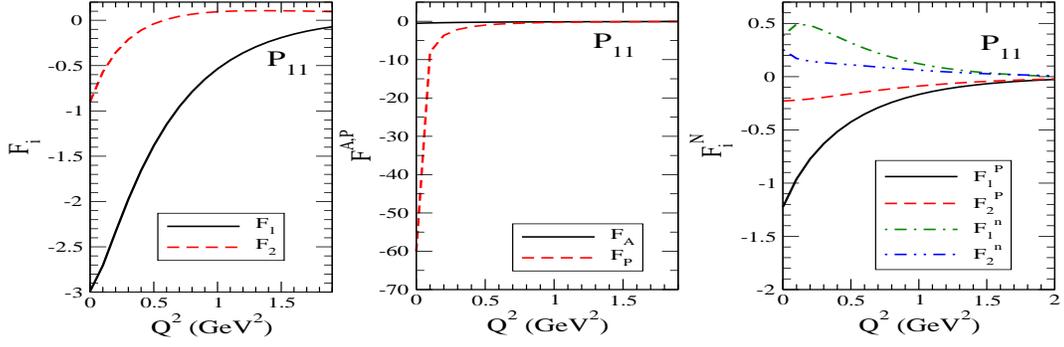}
\caption{$Q^2$ dependence of different form factors of $P_{11}(1440)$ resonance. 
The lines have same meaning as of Fig.~\ref{fig:S11_FF}.}
\label{fig:P11_FF}
\end{figure}

While the pseudoscalar form factor $F_P(Q^2)$  is related to the axial form factor $F_A(Q^2)$ via PCAC and is given by
\begin{equation}\label{eq:fp_res_spinhalf}
F_{P}(Q^2) = \frac{(MM_{R}\pm M^{2})}{m_{\pi}^{2}+Q^{2}} F_A(Q^2)
\end{equation}
where $+(-)$ sign is for positive(negative) parity resonances.
The $Q^2$ dependence of the form factors thus obtained are shown in Figs.~\ref{fig:S11_FF} and \ref{fig:P11_FF},  
 respectively for $S_{11}(1650)$ and $P_{11}(1440)$ resonant states. 

In analogy with Eq.~\ref{eq:res_had_current}, the most general form of the hadronic currents 
for the s-channel(direct diagram) and u-channel(cross diagram) processes
where a resonant state $R^{\frac12}$ is produced and decays to a pion in the final state, are written as 
\begin{eqnarray}\label{eq:res1/2_had_current}
j^\mu\big|_{R}^{\frac12}&=& 
i \; a\;  \mathcal{C^{R}} 
  \bar u(\vec{p}\,') 
 \slashchar{k}_\pi\gamma_5\frac{\slashchar{p}+\slashchar{q}+M}{(p+q)^2-M^2+ i\epsilon}\Gamma^\mu_{\frac12}  
u(\vec{p}\,),\nonumber \\\nonumber \\
j^\mu\big|_{C R}^{\frac12} &=& 
i \; a\; \mathcal{C^{R}} 
  \bar u(\vec{p}\,') \Gamma^\mu_{\frac12}
\frac{\slashchar{p}'-\slashchar{q}+M}{(p'-q)^2-M^2+ i\epsilon} \slashchar{k}_\pi\gamma_5  u(\vec{p}\,),
\end{eqnarray}
where $a=\cos \theta_c$ for the charged current process and $a= 1$ for the neutral 
current process. $\mathcal{C^{R}}$ is a constant which includes the coupling strength, isopin factor involve
in ${\cal R} \rightarrow N\pi$ transition, etc. This has been tabulated in Table-\ref{tab:Coupling constant}. 
 In the next section, we are going to present the method adopted to determine ${\cal R} \rightarrow N\pi$ coupling strength.
 
\subsubsection{Couplings of the resonances}\label{coupling}
Due to the lack of experimental data their is large uncertainty associated with ${\cal R}N\pi$ coupling at the ${\cal R}\to N\pi$ vertex. 
 We have fixed ${\cal R}N\pi$ coupling using the data of branching ratio and decay 
 width of these resonances from PDG~\cite{Olive:2014pdg} and use
 the expression for the decay rate which is obtained by writing the most general form of $RN\pi$ Lagrangian,
\begin{align}\label{eq:spin12_lag}
 \mathcal{L}_{R_{\frac{1}{2}} N \pi} &= \frac{f_{R\frac12} }{m_\pi}\bar{\Psi}_{R_{\frac{1}{2}}} \; \Gamma^{\mu}_{\frac{1}{2}}  \;
  \partial_\mu  \phi^i T_i \,\Psi \\
 \label{eq:spin32_lag} 
 \mathcal{L}_{R_{\frac{3}{2}} N \pi} &= \frac{f_{R\frac32}}{m_\pi} \bar{\Psi}_{R_{\frac{3}{2}}} \; \Gamma^{\mu}_{\frac{3}{2}} \;
\partial^\mu \phi^i T_i \,\, \Psi  
\end{align}
where $f_{R\frac12, R\frac32}$ is the ${\cal R}N\pi$ coupling strength. $\Psi$ is the nucleon field and ${\Psi}_{R_{\frac{1}{2}}}$ and ${\Psi}_{R_{\frac{3}{2}}}$ 
are the fields associated with the resonances of spin $\frac12$ and spin $\frac32$, respectively. 
$\phi^i$ are the pionic field and $T_i$ are the isospin operator which is $T = \tau$ for isospin  $\frac12$ 
states and $T = T^\dagger$ for 
isospin $\frac32$ states\footnote{$\vec \tau $ and $T^\dagger$ are the isospin operator for  doublet and  quartet, respectively.}. 
The interaction vertex $\Gamma^{\mu}_{\frac{1}{2}}$ is $ \gamma^\mu \gamma^5$($\gamma^\mu$)  for resonances(spin $\frac12$) 
with  positive(negative) parity. 
Similarly, the interaction vertex $\Gamma^{\mu}_{\frac{3}{2}}$(spin $\frac32$)  is ${\mathbb I}_4 $ 
for positive parity state and $\gamma_5$ for
negative parity state. Using the above Lagrangian one may obtain the expression for the 
decay width in the resonance rest frame as
\begin{align}\label{eq:12_width}
 \Gamma_{R_{\frac{1}{2}} \rightarrow \pi N} &= \frac{\mathcal{C}}{4\pi} \left(\frac{f_{R\frac12}  }{m_\pi}\right)^2 \left(M_R \pm M\right)^2 
\frac{E_N \mp M}{M_R} |\vec{q}_{\mathrm{cm}}| \\ 
\Gamma_{R_{\frac{3}{2}} \rightarrow \pi N} &= \frac{\mathcal{C}}{12\pi} \left(\frac{f_{R\frac32}}{m_\pi}\right)^2 \frac{E_N \pm M}{M_R} |\vec q_{cm}|^3,
\end{align}

where the upper(lower) sign represents the positive(negative) parity resonant state. 
The parameter  $\mathcal{C}$ is obtained from the isospin analysis and found out to be  $3$ for isospin $\frac12$ state and 
$1$ for isospin $\frac32$ states. $|\vec q_{cm}|$ is the outgoing pion momentum measured from resonance rest frame and is given by, 
\begin{equation}
|\vec{q}_{\mathrm{cm}}| = \frac{\sqrt{(W^2-m_{\pi}^2-M^2)^2 - 4 m_{\pi}^2 M^2}}{2 M_R}  \label{eq:pi_mom}
\end{equation}
and $E_N$, the nucleon energy is
\begin{equation}
  E_N=\frac{W^2+M^2-m_{\pi}^2}{2 M_R},
\end{equation}
where $W$ is the total center of mass energy carried by the resonance. 
In view of the above, we fix $N \Delta \pi$ coupling$(f_{\pi N \Delta})$ by comparing 
$\Delta \to N \pi$ decay width evaluated in the rest frame of $\Delta$, 
\begin{equation}
\Gamma_\Delta(s) = \frac{1}{6\pi} \left ( \frac{f_{\pi N \Delta}}{m_\pi}\right )^2
 \frac{M}{\sqrt s} \left [ \frac{\lambda^\frac12
  (s,m_\pi^2,M^2)}{2\sqrt s} \right ]^3 \Theta(\sqrt s
-M-m_\pi),\qquad s= p_\Delta^2 
\end{equation}
where  $\lambda(x,y,z) = x^2+y^2+z^2-2xy-2xz-2yz$ is 
K\"{a}llen function. To get the offshell effect of $\Delta(1232)$ resonance we have
taken momentum dependent width. 
\begin{figure}
\includegraphics[width=14cm,height=5.5cm]{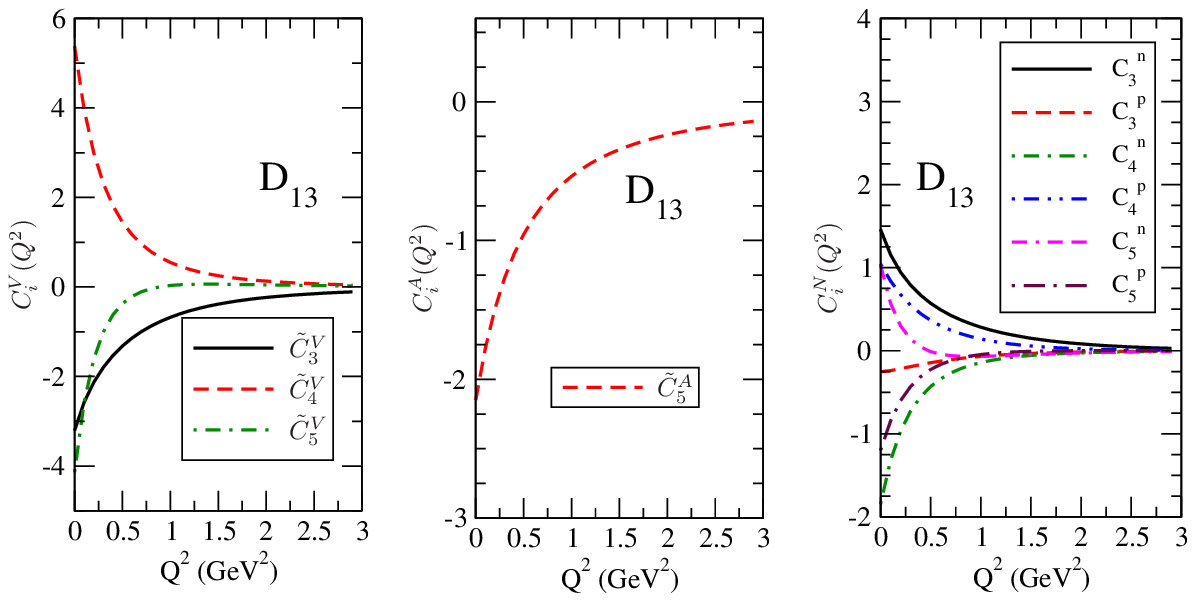}
\caption{$Q^2$ dependence of different form factors of $D_{13}$ resonance. From left to right panel: ${\tilde C}_3^V$, ${\tilde C}_4^V$ and ${\tilde C}_5^V$ 
mentioned in Eq.~\ref{eq:c3v}, and ${\tilde C}_5^A$ mentioned in Eq.\ref{c5a-r}, and $C_i^{n,p}, i=3,4,5$ mentioned in Eq.~\ref{eq:civ_NC}.}
\label{fig:D13_FF}
\end{figure}
Using the above expressions for decay width the 
coupling for $ {\cal R} \to N \pi$ are obtained and given in Table-\ref{tab:included_resonances}. 

\begin{table*}[t]
  \begin{center}
    \vspace{1cm}
    \begin{tabular*}{139mm}{@{\extracolsep{\fill}}c c c c c c c c c}
      \noalign{\vspace{-8pt}}
      \hline \hline
      Resonances               & $M_R$ [GeV] & J\quad   & I \quad    &   P   & $\Gamma_0^{tot}$  &  $\pi N$ branching  & $F_A(0)$ &   $f^\star$  \\
      &&&&&(GeV)&ratio ($\%$)  & or  ${\tilde C}_5^A(0)$\\ \hline
      $P_{33}$(1232)  &       $1.232$ & $3/2$ &$ 3/2$ &$ +$ &                $ 0.120$ &       $ 100$ &$ 1.0$ &   $2.14 $    \\ \hline
      
      $P_{11}$(1440)  &       $1.462$ & $1/2$ &$ 1/2$ &$ +$ &                $ 0.250$ &       $ 65$ &$ -0.43$&   $0.215 $	 \\ \hline
      
      $D_{13}$(1520)  &       $1.524$ & $3/2$ &$ 1/2$ &$ -$ &                $ 0.110$ &       $ 60$ &$-2.08$&  $1.575 $	 \\ \hline
      
      $S_{11}$(1535)  &       $1.534$ & $1/2$ &$ 1/2$ &$ -$ &                $ 0.151$ &       $ 51$ &$0.184$&	   $ 0.092 $	 \\ \hline
      
      $S_{11}$(1650)  &       $1.659$ & $1/2$ &$ 1/2$ &$ -$ &                $ 0.173$ &        $ 89$ &$-0.21$& $ - 0.105 $	 \\ \hline
      
      $P_{13}$(1720)  &       $1.717$ & $3/2$ &$ 1/2$ &$ +$ &                $ 0.200$ &       $ 11$ &$ -0.195$& $ 0.147 $	  \\ \hline
      \hline \hline
    \end{tabular*}
  \end{center}
  \caption{Properties of the 
  resonances included in the present model, with Breit-Wigner mass $M_R$, spin J, 
  isospin I, parity P, the total decay width $\Gamma_0^{tot}$, 
  the branching ratio into $\pi$ N, the axial coupling ($F_A(0)$ for 
  spin $\frac{1}{2}$ states; $C_5^A(0)$ for states with spin $\frac{3}{2}$) and $f^\star$ stands for $f_{R\frac12}$ or $f_{R\frac32}$ given in
  Eqs.\ref{eq:spin12_lag} and \ref{eq:spin32_lag} and for $\Delta(1232)$ resonance $f^\star=f_{\Delta N \pi}$.}
   \label{tab:included_resonances}
   \vspace{15mm}
\end{table*}
\begin{figure}
\includegraphics[width=14cm,height=5.5cm]{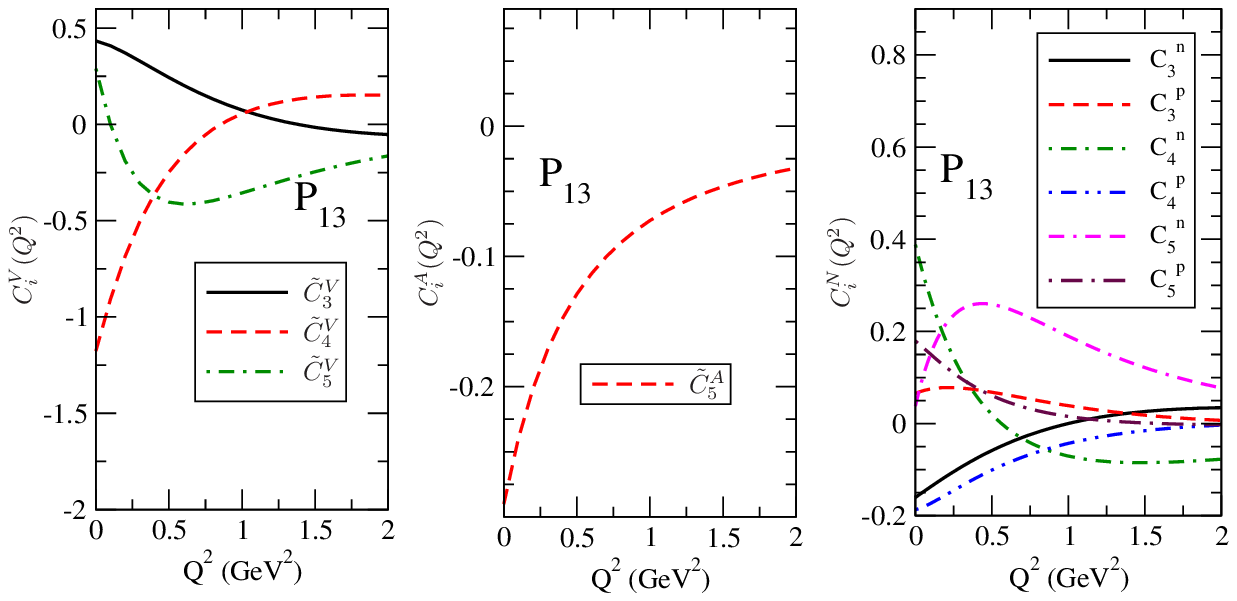}
\caption{$Q^2$ dependence of different form factors of $P_{13}$ resonance. From left to right panel: ${\tilde C}_3^V$, ${\tilde C}_4^V$ and ${\tilde C}_5^V$
as mentioned in Eq.~\ref{eq:c3v}, ${\tilde C}_5^A$ as mentioned in Eq.\ref{c5a-r} and $C_i^{n,p}, i=3,4,5$ mentioned in Eq.~\ref{eq:civ_NC}.}
\label{fig:P13_FF}
\end{figure}

\begin{table*}[h]
  \begin{center}
  \caption{Coupling constant($\mathcal{C^R}$) for spin $\frac12$ and spin $\frac32$ resonances.
Here $f^\star$ stands for ${\cal R} \to N \pi$ coupling which is for $\Delta(1232)$ resonance is $f_{\Delta N \pi}$
and $f_{R\frac12}(f_{R\frac32})$ for spin $\frac12(\frac32)$ resonances given in
  Eqs.\ref{eq:spin12_lag} and \ref{eq:spin32_lag}. }
  \label{tab:Coupling constant}
  \vspace{1cm}
    \renewcommand{\arraystretch}{1.5}
    \begin{tabular*}{139mm}{@{\extracolsep{\fill}}|c|ccc|ccc|}
      \noalign{\vspace{-8pt}}
      \hline \hline
      Process         & \multicolumn{3}{c|}{$\mathcal{C^R}$(CC $\nu$)}&\multicolumn{3}{c|}{$\mathcal{C^R}$(CC $\bar \nu$)}
           \\ \hline
      &   $p\pi^{+}$ & $n\pi^{+}$ & $p\pi^{0}$&
      $n\pi^{-}$ & $n\pi^{0}$ & $p\pi^{-}$
      \\ \hline
      $P_{33}$(1232)  &   $\frac{\sqrt{3}f^\star}{m_\pi}$   
      & $\sqrt{\frac{1}{3}}\frac{f^\star}{m_\pi}$& -$\sqrt{\frac{2}{3}}\frac{f^\star}{m_\pi}$ &
              $\frac{\sqrt{3}f^\star}{m_\pi}$  & $\sqrt{\frac{2}{3}}\frac{f^\star}{m_\pi}$& $\sqrt{\frac{1}{3}}\frac{f^\star}{m_\pi}$ \\
      $P_{11}$(1440) &   0  &-$\sqrt{\frac{1}{2}}\frac{f^\star}{f_\pi}$ & -$\frac{1}{2}\frac{f^\star}{f_\pi}$  &
              0 & -$\frac{1}{2}\frac{f^\star}{f_\pi}$ & $\sqrt{\frac{1}{2}}\frac{f^\star}{f_\pi}$ \\
      $D_{13}$(1520)  &   0 &-$\sqrt{\frac{1}{2}}\frac{f^\star}{m_\pi}$ & -$\frac{1}{2}\frac{f^\star}{m_\pi}$   &
              0 & -$\frac{1}{2}\frac{f^\star}{m_\pi}$ & $\sqrt{\frac{1}{2}}\frac{f^\star}{m_\pi}$ \\
      $S_{11}$(1535)  &   0 &-$\sqrt{\frac{1}{2}} \frac{f^\star}{f_\pi} $ & -$\frac{1}{2} \frac{f^\star}{f_\pi} $   &
              0 & -$\frac{1}{2} \frac{f^\star}{f_\pi} $ & $\sqrt{\frac{1}{2}} \frac{f^\star}{f_\pi} $ \\
      $S_{11}$(1650)  &   0 &-$\sqrt{\frac{1}{2}} \frac{f^\star}{f_\pi} $ & -$\frac{1}{2} \frac{f^\star}{f_\pi} $   &
              0 & -$\frac{1}{2} \frac{f^\star}{f_\pi} $ & $\sqrt{\frac{1}{2}} \frac{f^\star}{f_\pi} $ \\
      $P_{13}$(1720)  &   0 &-$\sqrt{\frac{1}{2}}\frac{f^\star}{m_\pi}$ & -$\frac{1}{2}\frac{f^\star}{m_\pi}$   &
              0 & -$\frac{1}{2}\frac{f^\star}{m_\pi}$ & $\sqrt{\frac{1}{2}}\frac{f^\star}{m_\pi}$ \\
      \hline \hline
    \end{tabular*}
  \end{center}
\end{table*}
\begin{table}
\begin{center}
\caption{ \label{tab:param-p1} MAID2008 parameterization of the transition
form factors for proton target. $\bar{\mathcal A}_{\alpha}(0)$ is given in 
units of $10^{-3}\,{\rm {GeV}}^{-\frac{1}{2}}$ and the
coefficients $a_1,\,a_2,\,a_4,\,b_1$ in units of ${\rm {GeV}}^{-2},\,{\rm
{GeV}}^{-4},\,{\rm {GeV}}^{-8},\,{\rm {GeV}}^{-2}$, respectively. For all fits
$a_3=0$.}
\begin{tabular*}{139mm}{@{\extracolsep{\fill}}ccccccc}
\hline
 $N^*,\;\Delta^*$ & Amplitude & $\bar{\mathcal A}_{\alpha}(0)$ &
$a_1$ & $a_2$ & $a_4$ & $b_1$ \\
\hline
$P_{11}(1440)$ & $A_\frac{1}{2}$ & $-61.4$ & $ 0.871$ & $-3.516$ & $-0.158$ & $1.36$\\
                & $S_\frac{1}{2}$ & $  4.2$ & $  40.0 $ & $  0   $ & $ 1.50 $ & $1.75$\\
$D_{13}(1520)$ & $A_\frac{1}{2}$ & $-27.4$ & $ 8.580$ & $-0.252$ & $ 0.357$ & $1.20$\\
                & $A_\frac{3}{2}$ & $160.6$ & $-0.820$ & $ 0.541$ & $-0.016$ & $1.06$\\
                & $S_\frac{1}{2}$ & $-63.5$ & $ 4.19 $ & $  0   $ & $   0  $ & $3.40$\\                
$P_{13}(1720)$ & $A_{\frac{1}{2}}$ & $ 73.0$ & $ 1.89 $ & $  0   $ & $   0  $ & $1.55$\\
                & $A_{\frac{3}{2}}$ & $-11.5$ & $10.83 $ & $-0.66 $ & $   0  $ & $0.43$\\
                & $S_{\frac{1}{2}}$ & $-53.0$ & $ 2.46 $ & $  0   $ & $   0  $ & $1.55$\\ \hline
\end{tabular*}
\end{center}
\end{table}
\begin{table}
\begin{center}
\caption{ \label{tab:param-p2} Maid2007 parameterization, Eq.~(\ref{eq:ffpar}),
for proton target ($a_{2,3,4}=0)$.}
\begin{tabular*}{139mm}{@{\extracolsep{\fill}}ccccc}
\hline
 $N^*,\;\Delta^*$ & Amplitude & $\bar{\mathcal A}_{\alpha}(0)$ & $a_1$ & $b_1$ \\
\hline
$S_{11}(1535)$ & $A_{\frac{1}{2}}$ & $ 66.4$ & $ 1.608$ & $0.70$\\
                & $S_{\frac{1}{2}}$ & $ -2.0$ & $  23.9$ & $0.81$\\
$S_{11}(1650)$ & $A_{\frac{1}{2}}$ & $ 33.3$ & $ 1.45 $ & $0.62$\\
                & $S_{\frac{1}{2}}$ & $ -3.5$ & $ 2.88 $ & $0.76$\\
\hline
\end{tabular*}
\end{center}
\end{table}

\vspace*{0.5cm}
\begin{table}
\begin{center}
\caption{ \label{tab:param-n}  Maid2007 parameterization for neutron target($a_{2,3,4}=0$).}
\begin{tabular*}{139mm}{@{\extracolsep{\fill}}ccccc}
\hline $N^*$ & Amplitude & $\bar{\mathcal A}_{\alpha}(0)$ & $a_1$ & $b_1$ \\
\hline
$P_{11}(1440)$ & $A_\frac{1}{2}$ & $ 54.1$ & $ 0.95 $ & $1.77$\\
                & $S_\frac{1}{2}$ & $-41.5$ & $ 2.98 $ & $1.55$\\
$D_{13}(1520)$ & $A_\frac{1}{2}$ & $-76.5$ & $-0.53 $ & $1.55$\\
                & $A_\frac{3}{2}$ & $-154.0$ & $ 0.58 $ & $1.75$\\
                & $S_\frac{1}{2}$ & $ 13.6$ & $ 15.7 $ & $1.57$\\
$S_{11}(1535)$ & $A_\frac{1}{2}$ & $-50.7$ & $ 4.75 $ & $1.69$\\
                & $S_\frac{1}{2}$ & $ 28.5$ & $ 0.36 $ & $1.55$\\
$S_{11}(1650)$ & $A_\frac{1}{2}$ & $  9.3$ & $ 0.13 $ & $1.55$\\
                & $S_\frac{1}{2}$ & $ 10. $ & $-0.5 $ & $1.55$\\
$P_{13}(1720)$ & $A_\frac{1}{2}$ & $ -2.9$ & $ 12.7$ & $1.55$\\
                & $A_\frac{3}{2}$ & $-31.0$ & $ 5.00 $ & $1.55$\\
                & $S_\frac{1}{2}$ & $  0  $ & $   0  $ & $  0 $\\ \hline
\end{tabular*}
\end{center}
\end{table}

\subsection{Neutral current neutrino(antineutrino) induced processes}\label{nc1pi}
In this section, we will briefly discuss the single pion production induced by neutral currents(NC1$\pi$).
Older data on NC1$\pi$ production are available from ANL~\cite{Derrick:1980nr} and Gargamelle~\cite{Pohl:1978iy} experiments.
Recently, NC1$\pi$ production measurements have been performed by the MiniBooNE~\cite{Anderson:2009zzc},
K2K~\cite{Nakayama:2004dp}, SciBooNE~\cite{Kurimoto:2009wq} collaborations. 
The neutral current $\pi^0$ production in neutrino interactions 
plays an important role in the background studies of
$\nu_\mu \leftrightarrow \nu_e$ or $\bar\nu_\mu \leftrightarrow \bar\nu_e$ oscillations in the appearance mode as well as in 
discriminating between $\nu_\mu \rightarrow \nu_\tau$ 
and $\nu_\mu \rightarrow \nu_s$ modes. Furthermore, in the reconstruction of neutrino energy using quasielastic events like
$\nu_e + n \rightarrow e^- + p$ or $\bar\nu_e + p \rightarrow e^+ + n$, 
a missing electron or positron in the $\pi^0$ decay may be mistaken as quasielastic event.  
 Moreover, neutral current induced pion production may also help 
 to distinguish between the production of $\nu_\tau$ and $\bar \nu_\tau$ 
in some oscillation scenarios at neutrino energies much below the $\tau$ production threshold but 
above the pion threshold.  

We have already discussed the contribution from non-resonant background terms for neutral current induced processes
in section~\ref{sec:nr_back}. Next we will present in brief the structure of 
resonant terms that may contribute to the hadronic current of (anti)neutrino induced neutral current processes. 
\subsubsection{Resonant contribution}
The other higher resonances like  $P_{11}(1440), D_{13}(1520), S_{11}(1535), S_{11}(1650)$ and $P_{13}(1720)$,  
 also contribute along with the dominant $\Delta(1232)$ resonance channel. 
The basic neutral current neutrino(antineutrino) induced reactions for pion production through resonance excitations are the following:
\begin{eqnarray}\label{eq:nc_res}
\nu_{_l}(k) + N(p) \rightarrow \nu_{l}(k^{\prime}) + {\cal R}(p_R) & \nonumber \\
 {\rotatebox[origin=c]{180}{{\huge$\Lsh$}}} \quad \; &
\hspace*{-0.65cm} N^\prime (p^\prime) + \pi (k_\pi)    
\end{eqnarray}
\begin{eqnarray}\label{eq:anc_res}
\bar\nu_{_l}(k) + N(p) \rightarrow \bar\nu_{l}(k^{\prime}) + {\cal R}(p_R) & \nonumber \\
 {\rotatebox[origin=c]{180}{{\huge$\Lsh$}}} \quad \; &
\hspace*{-0.65cm} N^\prime (p^\prime) + \pi (k_\pi)    
\end{eqnarray}
where ${\cal R}$ stands for the $\Delta(1232)$ and other higher resonances(R) which contribute to the pion production. 

In the next section, we will briefly discuss formalism to include different 
resonances for the charged current and the neutral current processes. \\

{{\bf A. Spin $\frac{3}{2}$ resonances}}\\

   The general structure for the hadronic current $J_{\mu}^{\frac{3}{2}}$ for neutral current induced 
   spin $\frac{3}{2}$ resonance in the intermediate state is given by Eq.~\ref{eq:had_current_3/2}, for which 
   $\Gamma_{\nu \mu}^{\frac32}$ is given by Eq.~\ref{eq:vec_3half_pos} for positive and negative parity states. 
   
 The vector and axial vector parts of the current(for $\Delta(1232), \; I = \frac32$ ) are given by Eq.~\ref{eq:vec_axial} with 
 the corresponding neutral current form factors ${({\tilde C}^V_i)^{NC}}~(i=3,4,5)$ and 
 ${({\tilde C}^A_i)^{NC}}~(i=4,5,6)$ which in the standard model are given 
 in terms of ${\tilde C}^{V}_i$ and ${\tilde C}^{A}_i$.  The expressions given in Eq.\ref{eq:vec_axial} now read as
  \begin{eqnarray}
{({\tilde C}^V_i)^{NC}}   &\rightarrow& (1-2 \sin^2 \theta_W) \tilde{C}^V_i, \\
{({\tilde C}^A_i)^{NC}}   &\rightarrow& - \tilde{C}^A_i.
\end{eqnarray}
Similarly for the case of $D_{13}(1520)$ and $P_{13}(1720)$, the neutral current form factors $\tilde{C}^V_i$ and $\tilde{C}^A_i$ are given by: 
\begin{align} 
{({\tilde C}^V_i)^{NC}} &\mathrel{\mathop{\longrightarrow}^{\mathrm{for \; p}}}  (1-2 \sin^2 \theta_W)  C^{p}_i  - \frac12   C^{n}_i     & i = 3,4,5         \\
{({\tilde C}^V_i)^{NC}} &\mathrel{\mathop{\longrightarrow}^{\mathrm{for \; n}}}  (1-2 \sin^2 \theta_W)  C^{n}_i  - \frac12   C^{p}_i     & i = 3,4,5         \\
{({\tilde C}^A_i)^{NC}} &\rightarrow  \pm  \frac12 \tilde{C}^A_i     & i = 5  
\end{align}
where plus(minus) sign stands for proton(neutron) targets. \\

{{\bf B. Spin $\frac{1}{2}$ resonances}}\\

For the neutral current process producing a  spin $\frac{1}{2}$ resonance in the intermediate state,
the hadronic current is given by 
Eq.~\ref{had_curr_1/2}. $\Gamma^\mu_{\frac12}$ is the vertex function
which for positive parity states is given by Eq.~\ref{eq:vec_half_pos} and for 
negative parity states is given by Eq.~\ref{eq:vec_half_neg}. 
The vector and axial vector parts of the current are written in terms of vector and axial vector
form factors and
have the same form as given in Eqs.~\ref{eq:vectorspinhalfcurrent} and \ref{eq:axialspinhalfcurrent}, 
but with a modified form factor corresponding to 
isospin $\frac{1}{2}$ resonance and a different expression for charged 
($\tilde{F}_i^p$) and neutral ($\tilde{F}_i^n$) resonant states with
the replacement of $F_{1,2}^{p,n}$  by $\tilde{F}_{1,2}^{p,n}$.

The explicit expressions for which are written as  

  \begin{eqnarray}
 \tilde{F}_i^p &=& (\frac{1}{2} -2 sin^{2}\theta_W) F_i^p - \frac{1}{2} F_i^n \nonumber \\
 \tilde{F}_A^p &=& \frac{1}{2} F_A 
 \end{eqnarray}
 for the positive charged state and 
 \begin{eqnarray}
  \tilde{F}_i^n&=&(\frac{1}{2} -2 sin^{2}\theta_W) F_i^n - \frac{1}{2} F_i^p \nonumber \\
  \tilde{F}_A^n&=&-\frac{1}{2} F_A
 \end{eqnarray}
 for the negative charged state. 
  $F_i$'s(i=1,2,A,P) are defined in section-\ref{spin12reso}.

\begin{center}
\begin{figure}
\includegraphics[width=12cm,height=10cm]{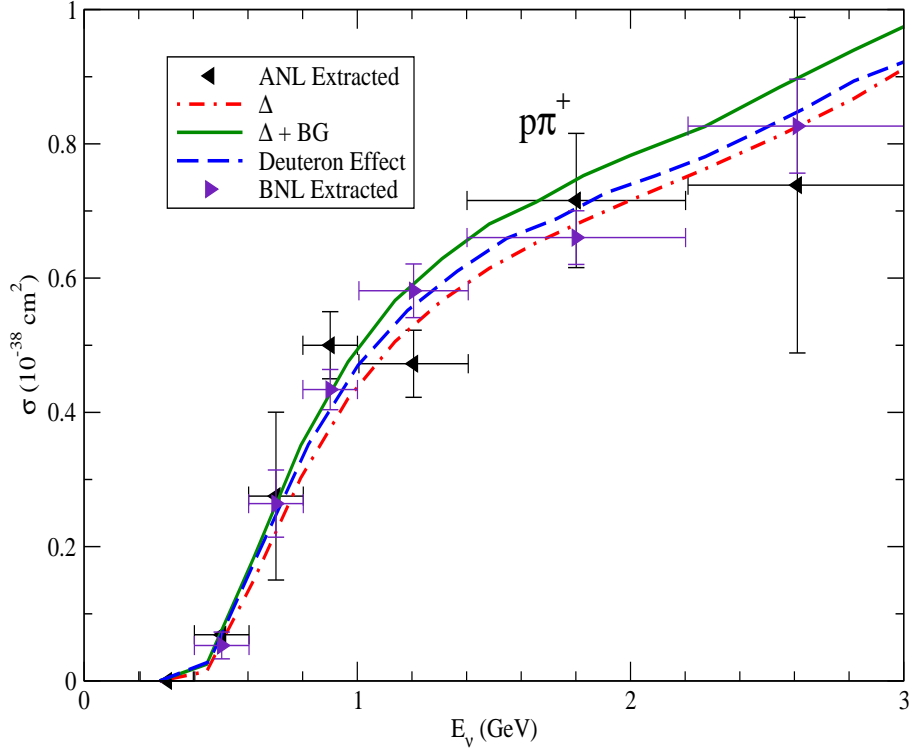}
\caption{Total scattering cross section for $\nu_{\mu}   p \rightarrow \mu^{-}   p   \pi^{+}$ process. 
 Experimental results are reanalyzed data points of ANL and BNL experiments by Wilkinson et al.~\cite{Wilkinson:2014yfa}.
No invariant mass cut has been applied. Dashed-dotted line is the result of the scattering cross section obtained by considering only the contribution from  
$\Delta(1232)$ resonance. When we also include 
non-resonant background terms in our calculations, 
the results are presented with solid line. The final results when deuteron effect is also taken into account has been shown by the long dashed line}
\label{fig:ANL_BNL}
\end{figure}
\end{center}
\begin{figure}
\includegraphics[width=14cm,height=13cm]{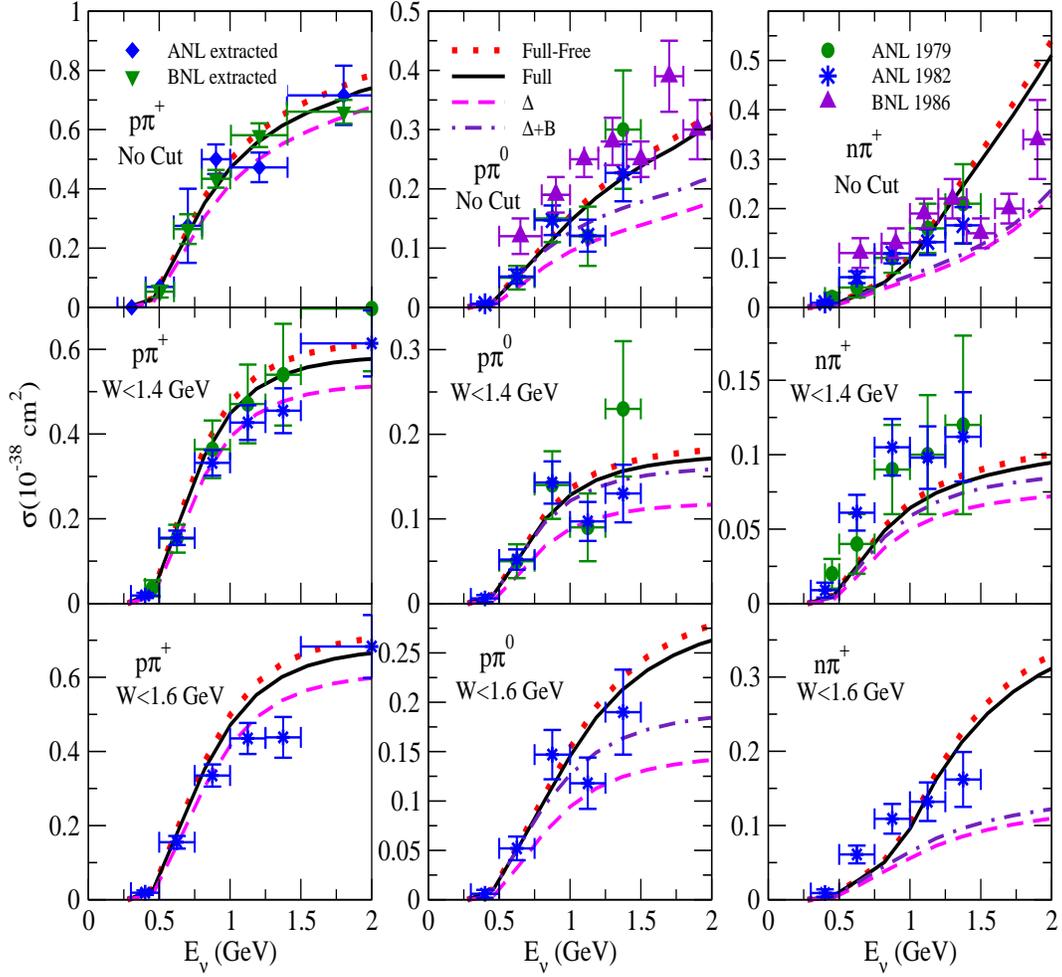}
\caption{Total scattering cross section for the charged current neutrino induced pion production processes: 
$\nu_{\mu}   p \rightarrow \mu^{-}   p   \pi^{+}$(Left panel), $\nu_{\mu}   n \rightarrow \mu^{-}   p   \pi^{0}$(Central panel), 
$\nu_{\mu}   n \rightarrow \mu^{-}   n   \pi^{+}$(Right panel).
 The dashed line is the result
calculated in the $\Delta(1232)$ dominance model, dashed-dotted line is the result obtained when we include non-resonant background terms in our calculations.
The solid line is the result of our full calculation when other resonances like
 $P_{11}(1440), D_{13}(1520), S_{11}(1535), S_{11}(1650)$  and $P_{13}(1720)$ are also included. All the above three cases are with deuteron effect.
 The dotted line is the result of the full calculation without deuteron effect.
The results in the top panels are obtained when we have not included any cut on invariant mass. 
The middle panel shows the results with a cut on center of mass energy of 1.4 GeV($W<1.4GeV$),
while in the bottom panel a cut of  $W<1.6GeV$ is introduced while calculating total scattering cross section. Data points quoted as ANL extracted
and BNL extracted are the reanalyzed data by Wilkinson et al.~\cite{Wilkinson:2014yfa}. Other data points in figures are the results 
from ANL~\cite{Radecky:1981fn} and BNL~\cite{Kitagaki:1986ct} experiments.}
\label{fig:sigma_Nu_CC1pion}
\end{figure}
\begin{figure}
\includegraphics[width=14cm,height=13cm]{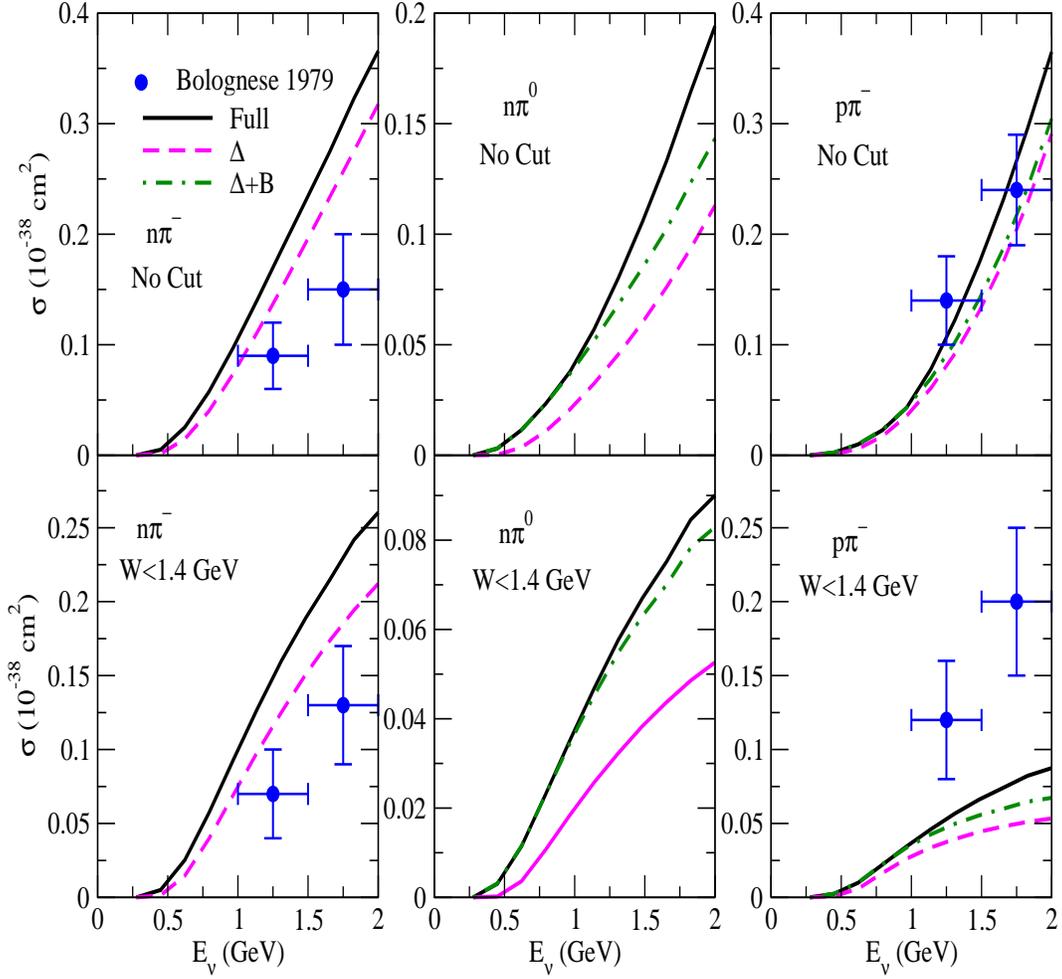}
\caption{Total scattering cross section for the charged current antineutrino induced pion production processes with deuteron effect.
 The results presented on the left panel are for $\bar\nu_{\mu}n \rightarrow \mu^{+}n\pi^{-}$, on the central panel are for 
 $\bar\nu_{\mu}  p \rightarrow \mu^{+}  
 n  \pi^{0}$, and on the right panel are for ~~$\bar\nu_{\mu}  p \rightarrow \mu^{+}  p  \pi^{-}$ processes.  
Data points are the experimental results from Ref.\cite{Bolognese:1979gf}. 
 The theoretical results presented here
    should be corrected for the nuclear medium effects before making any comparison with the experimental data.
Lines have the same meaning as in Fig.\ref{fig:sigma_Nu_CC1pion}.}
\label{fig:sigma_Nubar_CC1pion}
\end{figure}
\begin{figure}
\includegraphics[width=14cm,height=7cm]{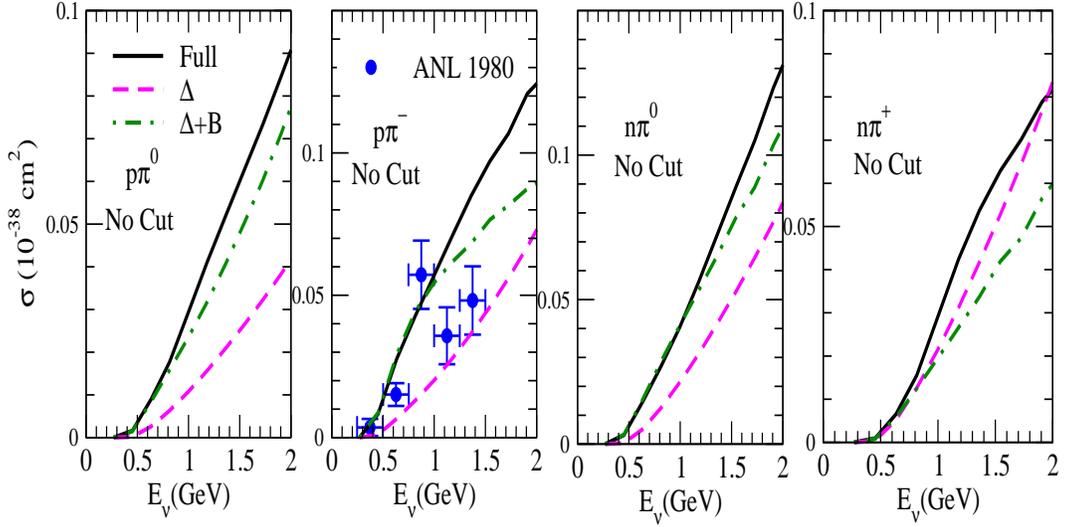}
\caption{Total scattering cross section for neutral current neutrino induced pion production processes  with deuteron effect. 
The results presented from the left to the right panels are for $\nu    p \rightarrow \nu   p   \pi^{0}$, 
 $\nu   n \rightarrow \nu    p   \pi^{-}$, $\nu    n \rightarrow \nu   n   \pi^{0}$,
 and $\nu    p \rightarrow \nu    n   \pi^{+}$ processes.
Data points are the experimental results from ANL~\cite{Derrick:1980nr} experiment. 
Results  are obtained without using any cut on the invariant mass W.
Lines have the same meaning as in Fig.\ref{fig:sigma_Nu_CC1pion}.}
\label{fig:sigma_Nu_NC1pion}
\end{figure}
\begin{figure}
\includegraphics[width=14cm,height=7cm]{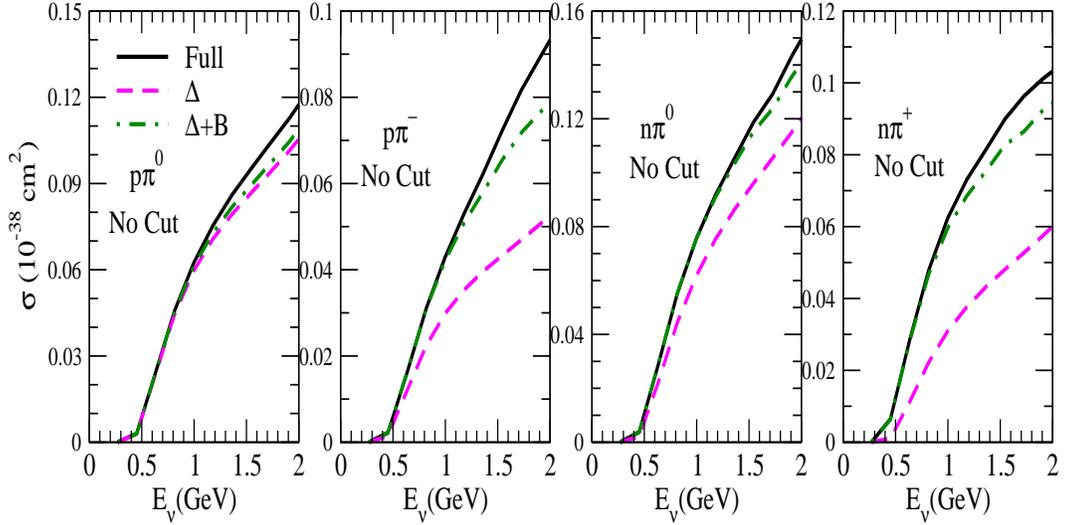}
\caption{Total scattering cross section for neutral current antineutrino induced pion production processes  with deuteron effect.
The results presented from the left to the right panels are for $\bar\nu   p \rightarrow \bar\nu   p  \pi^{0}$, 
 $\bar\nu   n \rightarrow \bar\nu   p  \pi^{-}$, $\bar\nu   n \rightarrow \bar\nu   n  \pi^{0}$, and 
 $\bar\nu   p \rightarrow \bar\nu   n  \pi^{+}$ processes. Results are obtained without using any cut on the invariant mass W.
Lines have the same meaning as in Fig.\ref{fig:sigma_Nu_CC1pion}.}
\label{fig:sigma_Nubar_NC1pion}
\end{figure}

\begin{figure}
\includegraphics[width=14cm,height=9cm]{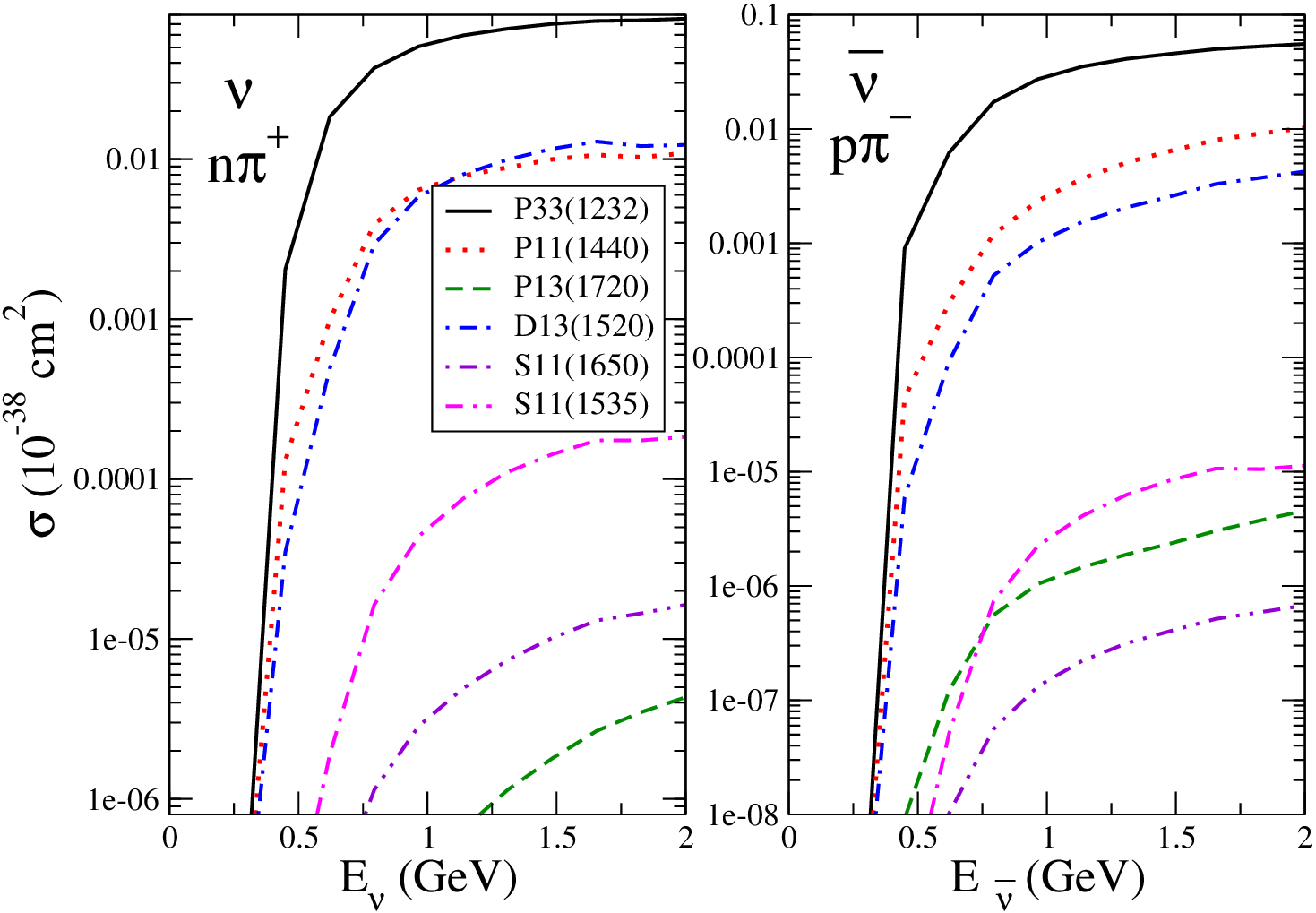}
\caption{The results are presented for the total scattering cross section for $\nu_{\mu}  n \rightarrow \mu^{-}  n  \pi^{+}$(Left panel) and 
 $\bar\nu_{\mu}  p \rightarrow \mu^{+}  p  \pi^{-}$(Right panel) processes where
 the individual contribution of various resonances have been shown.}
\label{fig:sigma_CC_res}
\end{figure}
\begin{figure}
\includegraphics[width=14cm,height=9cm]{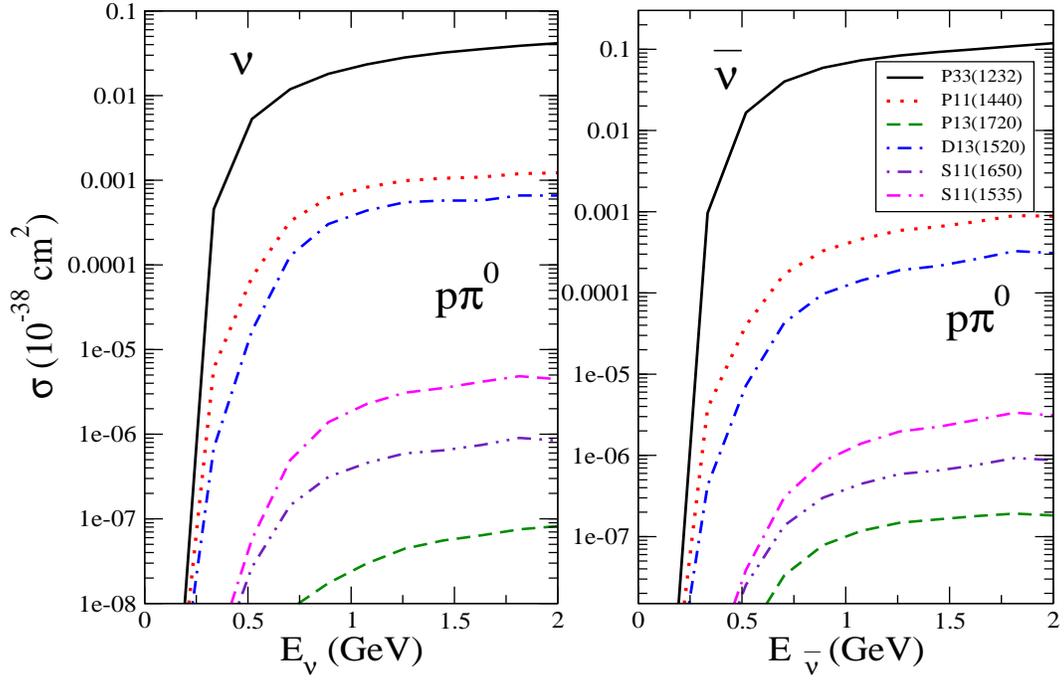}
\caption{The results are presented for the total scattering cross section for $\nu_{\mu}  p \rightarrow \nu_{\mu}  p  \pi^{0}$(Left panel) and 
 $\bar\nu_{\mu}  p \rightarrow \bar\nu_{\mu}  p  \pi^{0}$(Right panel) processes where
 the individual contribution of various resonances have been shown.}
\label{fig:sigma_NC_res}
\end{figure}
\begin{figure}
\includegraphics[width=12cm,height=8cm]{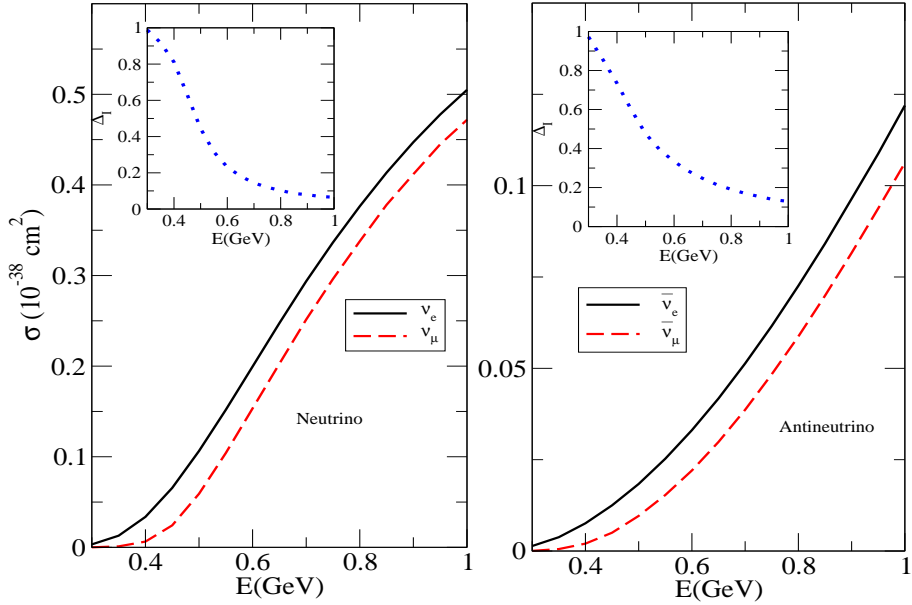}
\caption{Total scattering cross section for $\nu_{l}  p \rightarrow l^{-}  p  \pi^{+}$
and $\bar \nu_{l}  n \rightarrow l^{+}  n  \pi^{-}$ processes with deuteron effect where $l=e$(solid line), $\mu$(dashed line). In the inset the results for  
$\Delta_{I}= \frac{\sigma_{\nu_e(\bar\nu_e)}-\sigma_{\nu_\mu(\bar\nu_\mu)}}{\sigma_{\nu_e(\bar\nu_e)}} $ for neutrino(left panel)
 and antineutrino(right panel) induced processes have been shown. }
\label{fig:mass_effect}
\end{figure}
\section{Results and discussion}\label{results}
In this section, we present the results of the numerical calculations and discuss the findings. The results are presented for the 
total scattering cross section by integrating over the kinematical variables in Eq.\ref{eq:sigma_inelas}, where the matrix element is given by 
Eq.\ref{eq:Gg}. In the expression of the matrix element, the leptonic current is given by Eq.\ref{lep} and the hadronic current $j^{\mu\,{(H)}}$ is written as the sum of the contributions from the resonant terms including $\Delta(1232)$ resonance and non-resonant 
background(NRB) terms i.e. 
\begin{equation}\label{allterms}
 j^{\mu\,{(H)}}=j^{\mu}_{\Delta}~+~j^{\mu}_{\text NRB}~+~j^{\mu}_{\text R}, 
\end{equation}
where
\begin{eqnarray}\label{del-r}
 j^{\mu}_{\Delta}=j^{\mu}_{s, \Delta}~+~j^{\mu}_{u, \Delta} \nonumber\\
 j^{\mu}_{R}=j^{\mu}_{s, R}~+~j^{\mu}_{u, R} 
\end{eqnarray}
and $j^{\mu, CC}_{\text NRB}$ for the charged current induced processes is 
\begin{equation}\label{nrb_cc}
     j^{\mu, CC}_{\text NRB}=j^\mu\big|_{NP}~+~j^\mu\big|_{CP}~+~j^\mu\big|_{CT}~+~j^\mu\big|_{PP}~+~j^\mu\big|_{PF}, 
    \end{equation}
   and for the neutral current induced processes is
    \begin{equation}\label{nrb_nc}
     j^{\mu, NC}_{\text NRB}=j^\mu\big|_{NP}~+~j^\mu\big|_{CP} 
    \end{equation}
    
    In Eq.~\ref{allterms}, $j^{\mu}_{\text R}$ represents the contributions of all the higher resonances(other than $\Delta(1232)$) for the direct(s-channel) 
    and cross(u-channel) terms wherever applicable:
\begin{equation}\label{res-other}
     j^{\mu}_{\text R}= j^\mu_{P_{11}(1440)}~+~j^\mu_{S_{11}(1535)}~+~j^\mu_{S_{11}(1650)}~+~j^\mu_{D_{13}(1520)}~+~j^\mu_{P_{13}(1720)}. \nonumber
    \end{equation}
    
By $\Delta(1232)$ dominance we mean the contribution of s-channel as well as u-channel resonant terms (i.e. $=j^{\mu}_{s, \Delta}+j^{\mu}_{u, \Delta}$). 
$j^{\mu, CC}_{\text NRB}$ represents the hadronic current for charged current processes where the contributing terms are nucleon pole(NP), cross nucleon pole(CP),
contact term(CT), pion pole(PP) and pion in flight(PF) terms. 
$j^{\mu}_{\text R}$ denotes the hadronic current for the other higher resonances, for which we have taken the contribution from s- as well as u- channels.
    
 We must point out that in all the curves shown in the various figures, the places where we say $\Delta(1232)$ dominance means that the hadronic tensor
 $H^{\mu\nu}$ is obtained by evaluating
 $ {j^{\mu}_{s, \Delta}}^\dagger j^{\nu}_{s, \Delta} + {j^{\mu}_{s, \Delta}}^\dagger j^{\nu}_{u, \Delta} + {j^{\mu}_{u, \Delta}}^\dagger j^{\nu}_{s, \Delta} 
 + {j^{\mu}_{u, \Delta}}^\dagger j^{\nu}_{u, \Delta}$. When we say that the contributions from the background terms are also included, it is meant that the 
 hadronic tensor has now the contribution from the 
 square of $j^{\mu, CC}_{\text NRB}$ given in Eq.~\ref{nrb_cc} and the terms arising due to the 
 interference of ${j^{\mu}_{\Delta}}$ and $j^{\mu, CC}_{\text NRB}$ as given in Eqs.~\ref{del-r} and \ref{nrb_cc}, respectively.
 Finally, the results of full calculations would imply square of $j^{\mu\,{(H)}}$ given in Eq.\ref{allterms} to get the final expression for the hadronic tensor.
 
 Furthermore, this is to be pointed out that the non-resonant background terms have been obtained using SU(2)
   non-linear sigma Lagrangian for pions and nucleons interaction. Due to the limitations of this model at higher energies~\cite{Hernandez:2007qq}, we have 
   put a constraint on the center of mass energy(W) as $W_{min} = M + m_{\pi}$ and $W_{max} = 1.2 ~GeV$ while evaluating the non-resonant background terms. 
   We have varied $C_5^A(0)$ and axial mass $M_A$ in $C_5^A(Q^2)$ to get the 
   best description for the reanalyzed data~\cite{Wilkinson:2014yfa} of ANL and BNL experiments in the case of $\nu_\mu  p \rightarrow \mu^-  p  \pi^+$ process. 
      This constraint on W(i.e. $M + m_{\pi} \le W \le 1.2GeV$) has been put in all numerical evaluations while considering 
   non-resonant background contribution.
   
   Since earlier experiments to measure charged current neutrino induced single pion production were mainly performed using 
   hydrogen/deuteron target like the experiments at  ANL~\cite{Radecky:1981fn} and BNL~\cite{Kitagaki:1986ct}, therefore, deuteron correction
   factor must be taken into account. In a recent 
 analysis by Wilkinson et al.~\cite{Wilkinson:2014yfa} experimental results of ANL~\cite{Radecky:1981fn} and BNL~\cite{Kitagaki:1986ct} have 
 been normalized to deuteron data. Therefore, we have taken deuteron effect by following the prescription of Hernandez et al.~\cite{Hernandez:2010bx} and write
 \begin{equation} \label{de}
  \left(\frac{d\sigma}{dQ^2dW}\right)_{\nu d}=\int d{\bf p}_p^d |\Psi_d({\bf p}_p^d)|^2 \frac{M}{E_p^d} \left(\frac{d\sigma}{dQ^2dW}\right)_{\rm{ off~ shell}},
 \end{equation}
 where the four momentum of the proton inside the deuteron 
 is described by $p^\mu=(E_p^d, {\bf p}_p^d)$ with $E_p^d(=M_{\text Deuteron}-\sqrt{M^2+|{\bf p}_p^d|^2})$
 as the energy of the off shell proton 
 inside the deuteron and $M_{\text Deuteron}$ is the deuteron mass.
 $\left(\frac{d\sigma}{dQ^2dW}\right)_{\text off~shell}$ is obtained by using Eq.\ref{eq:sigma_inelas}.
  In the above expression $|\Psi_d|^2=|\Psi_0^d|^2~+~|\Psi_2^d|^2$, where $\Psi_0$ and $\Psi_2$ are the
  deuteron wave functions for the S-state and D-state, respectively and 
  have been taken from the works of Lacombe et al.~\cite{Lacombe:1981eg}.

  In Fig.~\ref{fig:ANL_BNL}, we have shown the results for the total scattering cross section for the 
  charged current neutrino induced 1$\pi^+$ production process on proton target i.e. for the reaction $\nu_\mu  p \rightarrow \mu^-  p  \pi^+$.
   In these calculations no invariant mass cut has been applied. The results are presented for the total scattering cross section in
   $\Delta(1232)$ - dominance model, then we include 
   the contributions from the non-resonant background(NRB) terms. 
   It may be pointed out that in the case of 1$\pi^+$ production process on proton target there is
   no contribution from the higher resonances considered here other than $\Delta(1232)$ resonance.
     The final results are with deuteron effect which has been obtained by using
   Eq.~\ref{de}. It may be observed that the inclusion of deuteron effect results into an overall reduction of $\sim 4-6\%$ in the total 
   scattering cross section. The present results are compared with the reanalyzed experimental analysis of
   ANL~\cite{Radecky:1981fn} and BNL~\cite{Kitagaki:1986ct} data by Wilkinson et al.~\cite{Wilkinson:2014yfa}. 
   We found the best fit of the total scattering cross section $\sigma(\nu_\mu  p \rightarrow \mu^-  p  \pi^+)$
   when $C_{5}^{A}(0) = 1.0$ and $M_{A} = 1.026$ GeV are used in the expression.
    
We have also calculated total scattering cross section for the charged current neutrino induced pion production processes and the 
results are presented in Fig.~\ref{fig:sigma_Nu_CC1pion}. The experimental points  for $\pi^+ p$ channel is the reanalyzed data
by Wilkinson et al.~\cite{Wilkinson:2014yfa} of the ANL~\cite{Radecky:1981fn} and BNL~\cite{Kitagaki:1986ct} experiments. While for 
the other channels like $\pi^0 p$ and $\pi^+ n$ the data are of ANL~\cite{Radecky:1981fn} and BNL~\cite{Kitagaki:1986ct} experiments.
In the case of $\nu_\mu  p \rightarrow \mu^-  p  \pi^+$ induced reaction,
 the main contribution to the total scattering cross 
 section comes from the $\Delta(1232)$ resonance and there is no contribution from the higher resonances which are considered here. 
 We find that due to the presence of the non-resonant background terms there is an increase in the cross section which is about $12\%$ at $E_{\nu_\mu}$=1GeV
   which becomes $\sim 8\%$ at $E_{\nu_\mu}$=2GeV.
  
  For $\nu_\mu  n \rightarrow \mu^-  n  \pi^+$ as well as $\nu_\mu  n \rightarrow \mu^-  p  \pi^0$
   processes, there are contributions from the non-resonant background terms as well as other higher resonant terms besides the $\Delta(1232)$ - 
  dominance. The net contribution to the total pion production due to the presence of the non-resonant 
  background terms in $\nu_\mu  n \rightarrow \mu^-  n  \pi^+$ 
  reaction results in an increase in the cross section 
  of about $12\%$ at $E_{\nu_\mu}$=1GeV 
 which becomes $6\%$ at $E_{\nu_\mu}$=2GeV. When other higher resonances are also taken
 into account there is a further increase in the cross section by about 
  $40\%$ at $E_{\nu_\mu}$=1GeV which becomes $55\%$ at $E_{\nu_\mu}$=2GeV. While in the case of $\nu_\mu  n \rightarrow \mu^-  p  \pi^0$   
   due to the presence of the background terms the total increase is about $26\%$ at $E_{\nu_\mu}$=1GeV and  $18\%$ at $E_{\nu_\mu}$=2GeV.
   Due to the presence of other higher resonances there is a further increase of about $35\%$ at $E_{\nu_\mu}$=1GeV and  $40\%$ at $E_{\nu_\mu}$=2GeV.
  Thus, we find that the inclusion of higher resonant terms lead to a significant increase in the cross section for 
  $\nu_\mu  n \rightarrow \mu^-  n  \pi^+$ and $\nu_\mu  n \rightarrow \mu^-  p  \pi^0$ processes. 
  Furthermore, it may also be concluded from the above observations that contribution from non-resonant background
  terms decreases with the increase in
 neutrino energy, while the total scattering cross section
  increases when we also include other higher resonances in our calculations.

    When a cut of $W \le 1.4GeV$ or $W \le 1.6GeV$ on the center of mass energy is applied then due to the presence of the non-resonant background terms,
  the increase in the total scattering cross section
  at $E_{\nu_\mu}$=1GeV for  $\nu_\mu  p \rightarrow \mu^-  p  \pi^+$    is about $10\%$ . 
  For $\nu_\mu  n \rightarrow \mu^-  n  \pi^+$ 
  reaction this increase in the cross section 
  is about $14\%$ at $E_{\nu_\mu}$=1GeV which becomes $5\%$ at $E_{\nu_\mu}$=2GeV.
  When other higher resonances are also taken into account there is a further increase in the cross section 
  which is about $40\%$ at $E_{\nu_\mu}$=1GeV. However, some energy dependence is observed and at $E_{\nu_\mu}$=2GeV
 this increase is $\sim 55\%$ for $W \le 1.4GeV$ and $\sim 65\%$ for $W \le 1.6GeV$. 
  While in the case of $\nu_\mu  n \rightarrow \mu^-  p  \pi^0$   
   due to the presence of the non-resonant background terms the total increase in cross section is about $27\%$ at $E_{\nu_\mu}$=1GeV  
   for $W \le 1.4GeV$ or $W \le 1.6GeV$.  Due to the presence of other resonances there is a further increase of about 10$\%$ 
   at $E_{\nu_\mu}$=1GeV.

   In Fig.~\ref{fig:sigma_Nubar_CC1pion}, we have shown the results for the charged current antineutrino induced pion production processes.
   These results are presented in the $\Delta(1232)$ dominance model, including non-resonant background terms as well as with our full prescription
   given in Eq.~\ref{res-other}.   
   Here also in the case of $\bar\nu_\mu  n \rightarrow \mu^+  n  \pi^-$ reaction there is
   no contribution from the higher resonances other than $\Delta(1232)$ resonance. 
   The inclusion of non-resonant background terms increases the cross section by around $24\%$ at $E_{\nu_\mu}$=1GeV which becomes
   around $12\%$ at $E_{\nu_\mu}$=2GeV. For $\bar\nu_\mu  p \rightarrow \mu^+  n  \pi^0$ reaction, 
   inclusion of non-resonant background terms increases cross section by around $42\%$ at $E_{\nu_\mu}$=1GeV which becomes 
   $20\%$ at $E_{\nu_\mu}$=2GeV. When other higher resonances are included, the cross section further increases by 
   $\sim 45\%$ at $E_{\nu_\mu}$=1GeV which becomes $40\%$ at $E_{\nu_\mu}$=2GeV.
   In the case of $\bar\nu_\mu  p \rightarrow \mu^+  p  \pi^-$ reaction, when we perform calculations by including non-resonant background 
   terms then cross section gets enhanced by around
   $16\%$ at $E_{\nu_\mu}$=1GeV which becomes $4\%$ at $E_{\nu_\mu}$=2GeV. When other higher resonances are also included along with 
   non-resonant background 
   terms then cross section further increases by around
   $18\%$ at $E_{\nu_\mu}$=1GeV and $\sim 20\%$ at $E_{\nu_\mu}$=2GeV.
   
   We have compared the present results with the experimental data of Gargamelle experiment performed at CERN PS where propane was used as the nuclear target. 
   Since propane is a composite target therefore the cross sections would get modulated due to nuclear medium effects.    
   Thus, the theoretical results presented in Fig. \ref{fig:sigma_Nubar_CC1pion}
    should be corrected for the nuclear medium effects before making any comparison with the experimental data. 
    We would like to point out that, in our earlier works~\cite{Ahmad:2006cy, SajjadAthar:2009rc, SajjadAthar:2009rd}
    of charged and neutral current pion production in the
    $\Delta(1232)$ dominance model, we have observed that nuclear medium effect reduces the cross section significantly when 
    the calculations are performed for nuclear targets and this helps in explaining the experimental data.
      
In Fig.~\ref{fig:sigma_Nu_NC1pion} (Fig.~\ref{fig:sigma_Nubar_NC1pion}), we have plotted the total scattering cross section for neutral 
current neutrino(antineutrino) induced pion production processes on proton and neutron targets. The experimental points  
are the data from ANL experiment~\cite{Derrick:1980nr}. Here also it may be observed that besides $\Delta(1232)$ resonant term, there is significant
contribution from non-resonant background terms which results in an increase in the total scattering cross section in all the channels. 
  Specifically, the increase in the cross section at $E_\nu =1$GeV due to non-resonant background terms in neutrino induced processes is $\sim$45$\%$ 
 for $\nu  p \rightarrow \nu  n \pi^+$,  $\sim$15$\%$ for $\nu  p \rightarrow \nu  p  \pi^0$, 
 $\sim$82$\%$ for $\nu  n \rightarrow \nu  p  \pi^-$ and $\sim$48$\%$ for $\nu  n \rightarrow \nu  n  \pi^0$. 
  Similarly, in the case of antineutrino induced processes the enhancement in the cross section due to the presence of non-resonant background terms 
  is $\sim$4$\%$ for $\bar \nu  p \to \bar \nu  p  \pi^0$, $\sim$49$\%$ for $\bar \nu  p \to \bar \nu  n  \pi^+$, 
  $\sim$18$\%$ for $\bar \nu  n \to \bar \nu  n  \pi^0$ and $\sim$30$\%$ for $\bar \nu  n \to \bar \nu  p  \pi^-$.
  We also observe that when higher resonant terms are included, there is no appreciable change in the cross sections which is in contrast to the 
observations made in the charged current induced reactions. For example, at $E_{\nu, \bar{\nu}} =1$GeV this increase is almost negligible 
 for all the antineutrino induced processes on proton and neutron targets as well as neutrino induced processes on neutron target. 
 There is $\sim 15\%$ enhancement in the cross sections in $\nu p \to \nu  \pi^+  n $ and $\nu  p \to \nu \pi^-  p $
 processes when higher resonant terms are included. 

 To explicitly show the contribution of individual resonances to the total scattering cross section, in Fig.~\ref{fig:sigma_CC_res}, we have
 presented the results for $\nu_{\mu}  n \rightarrow \mu^{-}  n  \pi^{+}$ and $\bar\nu_{\mu}  p \rightarrow \mu^{+}  p  \pi^{-}$ processes.
 It may be observed that the dominant contribution comes from $\Delta(1232)$ resonance followed by $P_{11}(1440)$ and $D_{13}(1520)$ resonances.
 However, the contribution for neutrino and antineutrino induced processes are not alike, for example, larger  $\Delta(1232)$ dominance
 may be observed in the neutrino case than in the case of antineutrino induced processes. For the case of neutrino induced charged current processes, 
 at $E_{\nu} = 1 GeV$, the contribution to the total scattering cross section from $P_{11}(1440)$($D_{13}(1520)$) resonance is around 
 $10 \%$($12 \%$) as that of the contribution from $\Delta(1232)$ resonance. However, at $E_{\nu} = 2 GeV$ contribution of $P_{11}(1440)$($D_{13}(1520)$)
  resonance is around $14 \%$($16 \%$).
  For antineutrino induced charged current processes, at $E_{\nu} = 1 GeV$, the contribution to the total scattering cross section 
  from $P_{11}(1440)$($D_{13}(1520)$) resonance is around 
 $8 \%$($2 \%$) which becomes around $18 \%$($6 \%$) at $E_{\nu} = 2 GeV$  
 as that of the contribution from $\Delta(1232)$ resonance. 

 Similar study has also been made for the (anti)neutrino induced neutral current processes. To explicitly show the contributions of different resonant terms,
 in Fig.~\ref{fig:sigma_NC_res}, we have presented the results for $\nu  p \to \nu  p  \pi^0$ and $\bar \nu  p \to \bar \nu  p  \pi^0$
 reactions.
 In this case it may be observed that the dominant contribution is still from $\Delta(1232)$ resonance followed by $P_{11}(1440)$ and $D_{13}(1520)$ resonances.
 However, a larger  $\Delta(1232)$ dominance
 is observed for antineutrino induced processes as compared to neutrino induced reaction, 
 which is in contrast to the observation in the case of charged current induced processes. 
 For example, for the case of neutrino induced neutral current processes, 
 the contribution to the total scattering cross section from $P_{11}(1440)$ resonance is  
 $3\%$ as that of the contribution from $\Delta(1232)$ resonance  in the energy range  $E_{\nu} = 1-2 GeV$. However, the contribution from 
 $D_{13}(1520)$ resonance in the energy range $E_{\nu} = 1-2 GeV$ is around $2 \% $.  
  For antineutrino induced charged current processes, in the energy range $E_{\nu} = 1-2 GeV$, the contribution to the total scattering cross section 
  from $P_{11}(1440)$($D_{13}(1520)$) resonance is almost negligible
 to that of the contribution from $\Delta(1232)$ resonance. 
 
 In Fig.~\ref{fig:mass_effect}, we have shown the lepton mass effect for the $\nu_e(\nu_\mu)$
and $\bar\nu_e(\bar\nu_\mu)$ induced processes by considering the reactions $\nu_{l}  p \rightarrow l^{-}  p  \pi^{+}$
and $\bar \nu_{l}  n \rightarrow l^{+}  n  \pi^{-}$. In the inset of these figures we have also shown the fractional change in the cross sections 
$\Delta_{I}= \frac{\sigma_{\nu_e(\bar\nu_e)}-\sigma_{\nu_\mu(\bar\nu_\mu)}}{\sigma_{\nu_e(\bar\nu_e)}} $ for neutrino(left panel)
 and antineutrino(right panel) induced processes. These results are shown up to $E_{\nu(\bar\nu)} \le 1GeV$. As may be observed from these curves that there is 
 significant effect of lepton mass(electron vs muon) on the total scattering cross section. 
 This study may be helpful in the analysis of data of the experiments planned in the $\sim1 GeV$ 
 energy region looking for the signals of CP violation in the leptonic sector.
 
 \section{Conclusions}\label{con}
In this work, we have presented a study of weak charged and neutral current induced single pion production from nucleons. 
The results have been presented for the total scattering cross sections by including the contributions of $P_{33}(1232)$, 
$P_{11}(1440)$, $S_{11}(1535)$, $D_{13}(1520)$, $S_{11}(1650)$ and $P_{13}(1720)$ resonances and the non-resonant background terms. 
Resonant terms are obtained using phenomenological
Lagrangian while non-resonant background terms
have been obtained using a SU(2) non-linear sigma model for pion-nucleon interaction Lagrangian.\\

We find that:

 \begin{enumerate}
  \item The pions are produced predominantly through $P_{33}(1232)$ resonance formation in $\nu_\mu  p \rightarrow \mu^-  p  \pi^+$ channel.
  The best description of the reanalyzed experimental 
 data of ANL and BNL experiments by Wilkinson et al.~\cite{Wilkinson:2014yfa} for this channel is obtained 
 when we take $C_5^A(0)$=1.0 and $M_A$=1.026GeV for N$-\Delta$ axial vector transition current form factor $C_5^A(Q^2)$.

 \item The enhancement in the cross section due to the presence of 
 non-resonant background terms at $E_\nu$ =1(2) GeV is around 12(8)$\%$ for $\nu_\mu  p \rightarrow \mu^-  p  \pi^+$ process,  and 
  24(12)$\%$ for $\bar\nu_\mu  n \rightarrow \mu^+  n  \pi^-$ process. 
  It may be noted that the energy dependence of non-resonant contributions in neutrino vs antineutrino induced processes is different.
 Higher resonances considered in this work do not contribute to these channels.
 
\item  For $\nu_\mu   n \rightarrow \mu^-   p  \pi^0$ process, we find the 
  enhancement in cross section to be around 26$\%$ at $E_\nu$ =1GeV when non-resonant background terms are included.
  There is a further increase of $\sim$9$\%$ in the cross section when higher 
  resonances are taken into account. The non-resonant contributions in the case of $\nu_\mu  n \rightarrow \mu^-  n  \pi^+$ process 
  is around 14$\%$ which  becomes  42$\%$ when higher resonances are also included. 
   The ratio of the cross sections for $\nu_\mu   n \rightarrow \mu^-   p  \pi^0$ to $\nu_\mu   n \rightarrow \mu^-   n  \pi^+$ processes is found
   to be $\sim$1.7 when evaluated in the $\Delta(1232)$ 
 dominance model, which becomes $\sim$2 if non-resonant background terms are included, and 1.5 when other higher resonances are also taken into account. 

 \item In the case of $\bar\nu_\mu  p \rightarrow \mu^+  n \pi^0$ process the enhancement in the cross section due to the presence of 
 non-resonant background terms is $\sim$42$\%$ at $E_\nu$ =1 GeV, which in the case of $\bar\nu_\mu  p \rightarrow \mu^+  p \pi^-$ process
 is $\sim$16$\%$. The contribution of higher resonance terms is quite small($\sim$3$\%$) in both the channels.  
The ratio of the cross sections for the processes $\bar\nu_\mu  p \rightarrow \mu^+  n \pi^0$ and 
 $\bar\nu_\mu  p \rightarrow \mu^+  p \pi^-$ is found to be $\sim$0.6  when evaluated in the $\Delta(1232)$ dominance model which becomes 
 $\sim$0.8 if non-resonant background terms are also included. 
   
   \item Qualitatively, the results for the neutral current induced
 pion production processes are similar to the charged ones, however, quantitatively there are 
 differences in the relative contribution of various terms.
 
 (i) The contribution of non-resonant terms leads to an increase in cross section for all the channels
  involving proton as well as neutron targets. The maximum contribution due to
  non-resonant terms is for $\nu  n \rightarrow \nu  p \pi^-$ process 
  and the minimum contribution is for $\bar\nu   p \rightarrow \bar\nu  p  \pi^0$. 
  
  (ii) When higher resonant terms are included, there is no appreciable change in the cross sections which is in contrast to the 
observations made in the charged current induced reactions. At $E_{\nu, \bar{\nu}} =1$GeV this increase is almost negligible 
 for all the antineutrino induced processes on proton and neutron targets as well as neutrino induced processes on neutron target. 
 In the case of other neutrino induced processes like $\nu  p \to \nu  \pi^+  n $ and $\nu  p \to \nu \pi^-  p $
this increase in cross section due to the inclusion of higher resonant terms is about $15\%$.

 \end{enumerate}
  These results may be used as a benchmark calculations for weak 
 charged and neutral current induced one pion production processes from nucleons.
 The present model can be applied to study the pion production from nuclear targets. 
 This work is presently going on and will be reported elsewhere.
\section{Acknowledgments}
M. S. A. is thankful to Department of Science
and Technology(DST), Government of India for providing financial assistance under Grant No. SR/S2/HEP-18/2012. 
M. R. A. and S. C. are thankful to University Grant Commission for providing financial assistance under UGC - Start up grant(No. F.30-90/2015(BSR)).


%
%
%
%
%
%
%
%
%
%


%
%

\end{document}